\documentclass[12pt]{article}
\usepackage[margin=3cm]{geometry}

\usepackage{lmodern}
\usepackage[utf8]{inputenc} % allow utf-8 input
\usepackage[T1]{fontenc}    % use 8-bit T1 fonts
\usepackage{hyperref}       % hyperlinks
\usepackage{url}            % simple URL typesetting
\usepackage{booktabs}       % professional-quality tables
\usepackage{amsfonts}       % blackboard math symbols
\usepackage{nicefrac}       % compact symbols for 1/2, etc.
\usepackage{microtype}      % microtypography

\usepackage{multirow}
\usepackage{amsmath}
\usepackage{amssymb}          %loads amsmath as well
\usepackage{graphicx}				% Use pdf, png, jpg, or eps§ with pdflatex; use eps in DVI mode
\usepackage{psfrag,epsf}
\usepackage{epstopdf}								% TeX will
\usepackage{siunitx}
\usepackage{caption}
\usepackage{float}
\usepackage{mathtools}
\usepackage{enumitem}
\usepackage{algorithmic}
\usepackage{algorithm}
\usepackage{authblk}
\hypersetup{colorlinks=true,linkcolor=blue,citecolor=blue,urlcolor=blue}
\usepackage{natbib}
\usepackage{subcaption}
\usepackage{subfiles}
\usepackage{authblk}

\newcommand{\phimat}{\bs \Phi}
\newcommand{\tmat}{{\bf T}}

\newcommand{\bbP}{\mathbb{P}}
\newcommand{\con}{{\,|\,}}
\newcommand{\T}{\textsc{T}}
\newcommand{\bs}[1]{{\boldsymbol{#1}}}

\newcommand{\multinomial}{\text{Multinomial}}
\newcommand{\bbI}{\mathbb{I}}
\newcommand{\betaprior}{\text{Beta}}
\newcommand{\dirichlet}{\text{Dir}}

\newtheorem{theorem}{Theorem}[section]
\newtheorem{proposition}{Proposition}[section]

\begin{document}

%\bibliographystyle{natbib}

% \def\spacingset#1{\renewcommand{\baselinestretch}%
% {#1}\small\normalsize} \spacingset{1}

%%%%%%%%%%%%%%%%%%%%%%%%%%%%%%%%%%%%%%%%%%%%%%%%%%%%%%%%%%%%%%%%%%%%%%%%%%%%%%

% \if1\blind
% {
%   \title{\bf Modeling Structure and Country-specific Heterogeneity in Misclassification Matrices of Verbal Autopsy-based Cause of Death Classifiers}
%   % \author{Sandipan Pramanik \hspace{.2cm}\\
%   %   Department of Biostatistics, Johns Hopkins University,\\
%   %   Scott Zeger \\
%   %   Department of Biostatistics, Johns Hopkins University\\
%   %   others \\
%   %   Abhirup Datta \\
%   %   Department of Biostatistics, Johns Hopkins University\\}
%   \author[1]{Sandipan Pramanik}
%   \author[1]{Scott Zeger}
%   \author[2]{Dianna Blau}
%   \author[1]{Abhirup Datta\footnote{Email address for correspondence: \href{mailto:abhidatta@jhu.edu}{abhidatta@jhu.edu}}}
%   \affil[1]{Department of Biostatistics, Johns Hopkins University}
%   \affil[2]{Global Health Center, US Centers for Disease Control and Prevention}
%   % \affil[2]{\pink{AD: Dianna, please add your affiliation here.}}
%   \maketitle
% } \fi

% \if0\blind
% {
%   \bigskip
%   \bigskip
%   \bigskip
%   \begin{center}
%     {\LARGE\bf Modeling Structure and Country-specific Heterogeneity in Misclassification Matrices of Verbal Autopsy-based Cause of Death Classifiers}
% \end{center}
%   \medskip
% } \fi

\title{\bf Modeling Structure and Country-specific Heterogeneity in Misclassification Matrices of Verbal Autopsy-based Cause of Death Classifiers}
  % \author{Sandipan Pramanik \hspace{.2cm}\\
  %   Department of Biostatistics, Johns Hopkins University,\\
  %   Scott Zeger \\
  %   Department of Biostatistics, Johns Hopkins University\\
  %   others \\
  %   Abhirup Datta \\
  %   Department of Biostatistics, Johns Hopkins University\\}
  \author[1]{Sandipan Pramanik}
  \author[1]{Scott Zeger}
  \author[2]{Dianna Blau}
  \author[1]{Abhirup Datta\footnote{Email address for correspondence: \href{mailto:abhidatta@jhu.edu}{abhidatta@jhu.edu}}}
  \affil[1]{Department of Biostatistics, Johns Hopkins University}
  \affil[2]{Global Health Center, US Centers for Disease Control and Prevention}
  % \affil[2]{\pink{AD: Dianna, please add your affiliation here.}}
\maketitle

\begin{abstract}
\textit{Verbal autopsy (VA) algorithms} are routinely used to determine individual-level \textit{causes of death (COD)} in many low-and-middle-income countries, which are then aggregated to derive population-level \textit{cause-specific mortality fractions (CSMF)}, essential to informing public health policies. However, VA algorithms frequently misclassify COD and introduce \textit{bias} in CSMF estimates. A recent method, \textit{VA-calibration}, can correct for this bias using a  VA misclassification matrix estimated from paired data on COD from both VA and {\em minimally invasive tissue sampling (MITS)} from the \textit{Child Health and Mortality Prevention Surveillance (CHAMPS) Network}. Due to the limited sample size, CHAMPS data are pooled across all countries, implicitly assuming that the misclassification rates are homogeneous. 
	
In this research, we show that the VA misclassification matrices are substantially heterogeneous across countries, thereby biasing the VA-calibration. We develop a coherent framework for modeling country-specific misclassification matrices in data-scarce settings. We first introduce a novel \textit{base model} based on two latent mechanisms -- \textit{intrinsic accuracy} and \textit{systematic preference} to parsimoniously characterize misclassifications. We prove that they are identifiable from the data and manifest as a form of invariance in certain misclassification odds, a pattern evident in the CHAMPS data. Then we expand from this base model, adding higher complexity and country-specific heterogeneity via interpretable \textit{effect sizes}. \textit{Shrinkage priors} balance the \textit{bias-variance tradeoff} by adaptively favoring simpler models. We publish uncertainty-quantified estimates of VA misclassification rates for 6 countries. This effort broadens VA-calibration's future applicability and strengthens ongoing efforts of using VA for mortality surveillance.

% The text of your abstract. 200 or fewer words.
\end{abstract}

\noindent%
{\it Keywords:}  Bayesian, hierarchical modeling, verbal autopsy, minimally invasive tissue sampling, global health.
% \vfill

% \newpage
% \spacingset{1.9} % DON'T change the spacing!
\section{Introduction}\label{sec:Introduction}

Understanding a population's burden of diseases is essential to effective public health practice. This heavily relies on \textit{cause-specific mortality fractions (CSMF)} that measure the prevalence of different diseases by aggregating individual-level \textit{causes of death (COD)} data. However, complete and accurate COD information from a \textit{complete diagnostic autopsy (CDA)} is uncommon in many \textit{low and middle-income countries (LMICs)} \citep{Nichols2018}. Therefore, alternative approaches like \textit{minimally invasive tissue sampling (MITS)} and \textit{verbal autopsy (VA)} are being increasingly relied upon. For COD determination, MITS is highly accurate relative to \textit{full diagnostic autopsies} \citep{bassat2017,menendez2017}. However, due to its resource-intensive nature, it is predominantly conducted in select hospitals and is therefore not \textit{scalable} to an entire population. 

VA is a non-invasive procedure that systematically interviews household members and gathers a decedent's health history and symptoms \citep{mikkelsen2015,wang2017,adair2018}. The interview responses are then input into a \textit{computer-coded VA (CCVA) algorithms} (referred to here as VA algorithm) to assign a COD. Given the scalability afforded by automation, VA has emerged as the predominant source for determining COD in community settings and is routinely used to estimate CSMF \citep{Savigny2017,garenne2014,soleman2006,setel2005}. \textit{Countrywide Mortality Surveillance for Action (COMSA)} programs have recently demonstrated this where nationwide VA studies were conducted in Mozambique (COMSA-Mz) and Sierra Leone to generate mortality rates and COD data for all ages to monitor the Sustainable Development Goals for health \citep{Fiksel2023, Gilbert2023,Macicame2023,Marsh2022}. %\pink{AD: Cite COMSA-Mz and COMSA-SL papers.}

While VA is the only feasible tool for mortality surveillance at national scales, VA-predicted causes of death (VA-COD) suffer from substantial \textit{misclassification bias} when compared to MITS or physician's diagnosis \citep{clark2018,adair2018,datta2020,fiksel2022}. A naive aggregation of VA-COD inherits this bias and produces highly inaccurate CSMF estimates. %Naively aggregating predicted VA-COD to produce CSMF estimates inherits the COD bias. %The bias then carries over into the \textit{raw (uncalibrated)} estimates of CSMFs that are often used in practice. 
Recent research has introduced \textit{VA-calibration} to rectify this anticipated bias \citep{datta2020,fiksel2022,Fiksel2023}. VA-calibration utilizes a limited set of paired CODs predicted from both VA and a standard reference diagnosis, such as MITS, to understand the misclassification rates of VA algorithms. For $C$ causes of death, the \textit{misclassification matrix} (also referred to as the {\em confusion matrix}) is a $C\times C$ matrix whose $(i,j)^{th}$ entry is the \textit{conditional probability} that the VA algorithm predicts cause $j$ given the true (or reference) cause $i$. Thus, its diagonal elements are \textit{sensitivities} and off-diagonal entries are \textit{false positive rates}. In a \textit{hierarchical Bayesian setup}, VA-calibration uses this matrix to \textit{calibrate} the raw estimates and it has been shown to significantly enhance the CSMF estimation accuracy. 

Recently, VA-calibration was used to produce CSMF estimates for neonatal and children (1--59 months) deaths at the national level in Mozambique \citep{Fiksel2023,Gilbert2023,Macicame2023}. 
%\pink{(AD: Also cite the Gilbert 2023 and Ivalda 2023 papers)}. 
Utilizing data collected in the \textit{Child Health and Mortality Prevention Surveillance (CHAMPS) project}, this study presents misclassification rate estimates of VA compared to MITS. CHAMPS records paired CODs predicted from both VA and MITS and currently spans six countries, including Mozambique.
%Using the data collected in {\em Child Health and Mortality Prevention Surveillance (CHAMPS)} project, VA misclassification rate estimates with respect to MITS which contained paired CODs predicted from VA and MITS across six countries including Mozambique. 
In order to enhance the sample size and \textit{precision} of estimates, the analysis \textit{pooled} CHAMPS data from all countries rather than solely relying on the data collected in Mozambique.
% To increase the sample size and improve the precision of estimates, the analysis \textit{pooled} CHAMPS data from all countries as opposed to using the data collected only in Mozambique.
%CHAMPS data were \textit{pooled} across all countries as opposed to using just the subset of data from Mozambique to increase the sample size of the paired data and improve the \textit{precision} of the estimates of the misclassification matrix.

The misclassification matrix is the cornerstone of VA-calibration and it is crucial to ensure the validity of its estimation. Pooling data from all countries implicitly makes a {\em homogeneity assumption} that misclassification rates are the same for all countries. In the context of VA-calibration, this represents a \textit{bias-variance tradeoff}. Thus any possible gains in precision from pooling the data are offset by possible bias introduced if data from other countries are not representative of Mozambique or other target countries. 

There are two statistical challenges in the estimation of VA misclassification matrices. First, in Figure~\ref{fig: observed classification proportions} we demonstrate that the observed misclassification proportions vary significantly across countries. The presence of \textit{country-specific heterogeneity} contradicts the assumption of homogeneity that was previously used to justify a pooled misclassification matrix. Consequently, the pooled estimates can introduce bias in the resulting CSMF estimates. Second, while it would be preferable to estimate VA-misclassification rates separately for each country to accommodate heterogeneity, the limited number of observed MITS cases per country presents a challenge, with  6 or fewer samples for 75\% of the cells in the matrix and 32\% of the cells have zero samples (See Figure~S1 in the supplement). Although this could potentially reduce bias (improve accuracy), the insufficient sample size would increase \textit{uncertainty} (reduce precision). This is undesirable for subsequent calibration and would also preclude extrapolation to countries that are not among the CHAMPS sites.

In this article, we propose a framework for modeling heterogeneity in VA misclassification matrices across countries with three key contributions that balance the bias-variance tradeoff. First, we introduce a novel and parsimonious {\em base model} for misclassification matrices that assumes a lower-dimensional structure based upon two novel mechanisms for correct or mis-classification -- {\em intrinsic accuracy} of a classifier to correctly identify the true cause, and the {\em systematic preference (or pull)} of the classifier towards predicting a cause regardless of what the true cause is. We show that these mechanisms, if present, can be identified based on observed misclassification counts of a classifier (See Theorem \ref{thm: logodds characterization}). Figure~\ref{fig: logodds} shows that both mechanisms are present for \textit{InsilicoVA}, the leading VA algorithm. The mechanisms borrow information across matrix cells leading to considerable dimension reduction, which is particularly advantageous in data-scarce settings.

The second contribution builds upon the parsimonious base model and proposes a nested hierarchical model for country-specific misclassification matrices with increasing complexity. This introduces a broader class of distributions on country-specific misclassification matrices that shrinks towards the base model. Each hierarchy promotes shrinkage towards the lesser complex model through three effect size parameters that explicitly control the \textit{pull strength} (informing on the degree of diversion from the base model) and the \textit{degree of heterogeneity}. We assume \textit{shrinkage priors} on them to promote \textit{continuous shrinkage}. In the absence of compelling evidence supporting complexity, the nested structure adaptively favors simpler models, enhancing the precision of estimates when dealing with limited sample sizes. We assessed the framework through extensive simulation studies and applied it to generate misclassification rates for InSilicoVA in six countries -- Bangladesh, Ethiopia, Kenya, Mali, Mozambique, and Sierra Leone. The estimates will be used to calibrate VA data for the next set of regional and national under-5 cause-specific mortality estimates for the Sustainable Development Goals \citep{perin2022global}.

%The degree of diversions from the base model and the homogenous model are capture with explicit dispersion parameters which are estimated in a data-driven manner, thereby adaptively adjusting the level of complexity needed for accurate modeling of the country-specific misclassification rates. 

%base model for misclassification  to mechanisms for correct or mis-classification of causes the concept of \textit{pull}, which quantifies \textit{systematic algorithmic preferences} and enables a structured modeling of misclassification matrix with considerably reduced complexity. Secondly, we define a hierarchy of models in a nested manner that expands from a simpler model under the assumption of a \textit{strong pull} to a more complex, country-specific model. Driven by data, it adaptively adjusts the level of complexity needed for accurate modeling. 
%Lastly, we introduce effect sizes that explicitly control the \textit{pull strength} and the \textit{degrees of heterogeneity}, and assume \textit{shrinkage priors} on them to promote \textit{continuous shrinkage}. This enables the framework to adaptively favor simpler models in the absence of substantial evidence, improving the stability of estimates in the presence of limited sample sizes.

The rest of the article is organized as follows. We describe the motivating dataset in Section~\ref{sec: Motivating Dataset From Child Health and Mortality Prevention Surveillance (CHAMPS) Network}. %, and Section~\ref{sec: Prior Research on VA-calibration} discusses the existing research.
In Section~\ref{sec: Method} we propose the methodology. Section~\ref{sec: Simulation Study} investigates its empirical properties through extensive simulation studies depicting the real-life application. We revisit the motivating dataset in Section~\ref{sec: Misclassification Analysis for InsilicoVA in CHAMPS} and apply the proposed framework to obtain the misclassification matrix estimate for \textit{InSilicoVA}, a widely used VA algorithm. We conclude in Section~\ref{sec: Discussion} with a discussion of findings and future research opportunities.

\section{Motivating Dataset From The \textit{CHAMPS Network}}\label{sec: Motivating Dataset From Child Health and Mortality Prevention Surveillance (CHAMPS) Network}

%\paragraph{Importance of \textit{CHAMPS} in verbal autopsy misclassification analysis.} 
%With the growing acceptance of VA, \textit{VA-calibration} has emphasized the importance of integrating misclassification rates \citep{datta2020,fiksel2022}. 

\textit{The Child Health and Mortality Prevention Surveillance (CHAMPS)} is an ongoing initiative dedicated to comprehensive child mortality surveillance in multiple countries \citep{blau2019,Salzberg2019}. For a limited number of deaths among neonates, children aged 1-59 months, and stillbirths in hospital facilities, CHAMPS performs \textit{minimally invasive tissue sampling (MITS)} to obtain a comprehensive COD diagnosis. With VA-predicted CODs also recorded for the deaths, this generates a labeled dataset of paired MITS- and VA-CODs across multiple sites in South Asia and Africa. This constitutes a unique resource for assessing the algorithmic precision of VA in estimating mortality rates.%The multi-country CHAMPS dataset of VA-MITS pairs constitutes a unique resource to evaluate the algorithmic accuracy of VA in estimating mortality rates. 

The previous analysis of pooled CHAMPS COD data using VA-calibration unveiled substantial misclassification rates across age groups for various VA algorithms \citep{Fiksel2023,Gilbert2023}. Subsequently, CSMF estimates for neonates and children aged 1-59 months were obtained in Mozambique by calibrating VA-only data from COMSA-Mz. %The \textit{calibrated CSMF} estimates from VA-calibration significantly differed from their raw estimates. 
%This has also been recently confirmed in the analysis of VA data from the \textit{Countrywide Mortality Surveillance for Action project} in Mozambique where the \textit{calibrated CSMF} estimates from VA-calibration significantly differed from their raw estimates \citep{fiksel2023}. 
Pooling of CHAMPS data across all the countries implicitly makes the assumption that the misclassification rates are homogeneous, and was done in \cite{Fiksel2023} to increase sample size and precision of the misclassification matrix estimate. With more data being collected, one can assess for heterogeneity in the misclassification rates across countries. 

\begin{figure}[!t] %  figure placement: here, top, bottom, or page
   \centering
   \includegraphics[width=.98\textwidth]{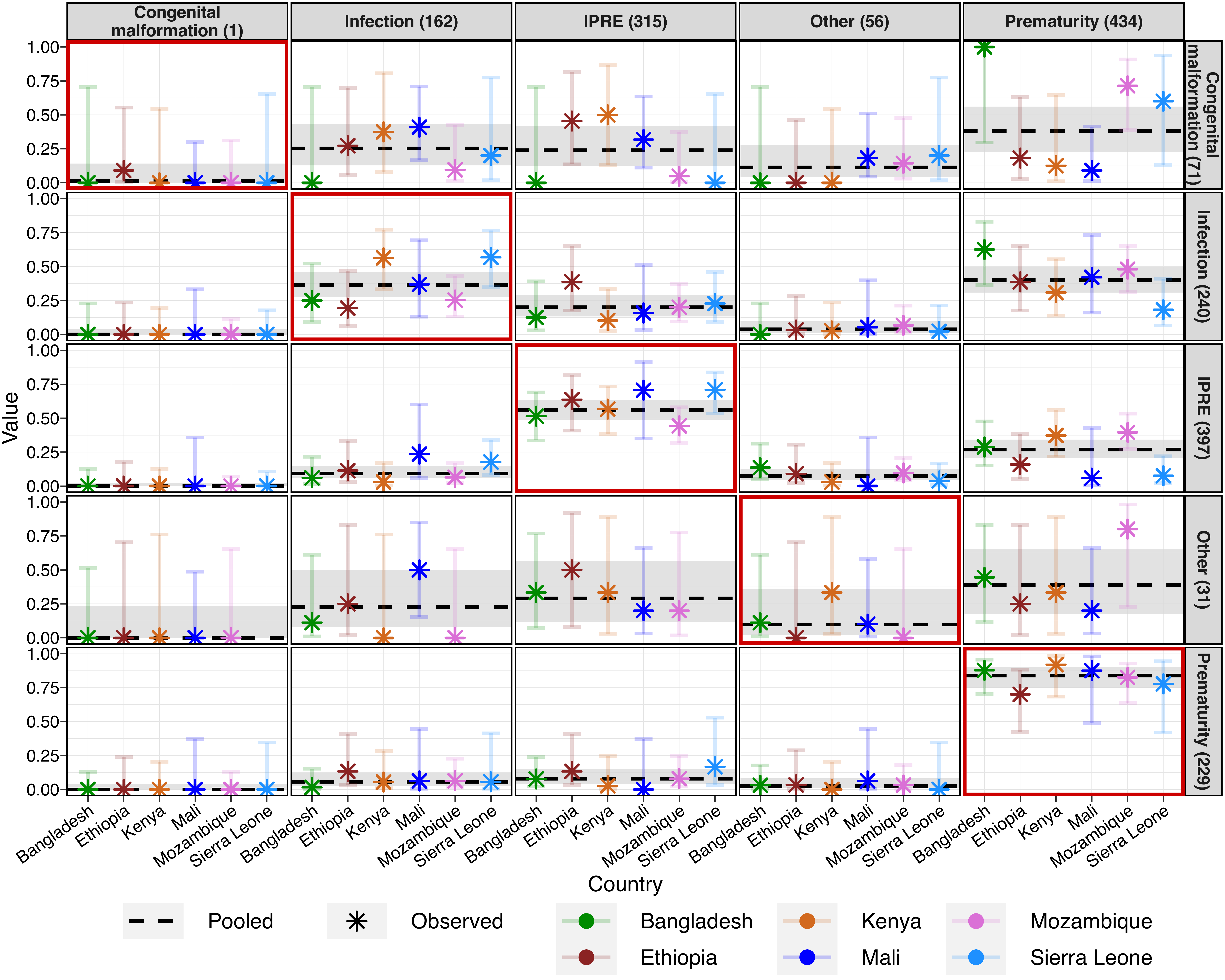}
   \caption{Country-specific empirical misclassification matrices of InSilicoVA concerning MITS for neonatal deaths in the CHAMPS dataset. Rows and columns are MITS and VA-predicted causes. The diagonal and off-diagonal panels respectively display the sensitivities and false positives. The dashed black horizontal lines are average rates, \textit{pooled} across countries. The error bars and grey bands are 95\% confidence intervals (obtained using \texttt{MultinomCI()} in \texttt{R} package \texttt{DescTools}). Numbers in parentheses denote the sample size.}
   \label{fig: observed classification proportions}
\end{figure}
In Figure \ref{fig: observed classification proportions}, we analyze the misclassification pattern for InSilicoVA using 570 neonatal deaths that occurred within CHAMPS network hospitals in Bangladesh, Ethiopia, Kenya, Mali, Mozambique, and Sierra Leone between July 2017 and 2023. %Our analysis further focuses on cases of \textit{single-cause} COD, where only the \textit{top-cause} is documented. 
Following \cite{Fiksel2023}, both VA and MITS-COD are aggregated into $C=5$ broad categories: \textit{congenital malformation}, \textit{infection}, \textit{intrapartum related events (IPRE)}, \textit{prematurity}, and \textit{other}. 

The Figure displays observed misclassification proportions in each country together with the pooled proportion (dashed black line). The figure highlights two main challenges. 
\begin{itemize}
    \item {\bf \em Heterogeneity across countries.} Compared to the pooled values, the country-specific proportions substantially deviate across countries for several MITS-VA cause pairs, indicating a significant variation in InSilicoVA's misclassification between countries. This is particularly pronounced for sensitivities of infection and IPRE, and false positive proportions for MITS-VA cause pairs congenital malformation-IPRE and infection-prematurity. Given the current implementation of VA-calibration is based on the homogeneity assumption, this would produce locally incongruous misclassification rate estimates and could adversely affect the accuracy of calibrated CSMF estimates.
    \item {\bf \em Limited size of CHAMPS data.} In the presence of a strong heterogeneity as in Figure \ref{fig: observed classification proportions}, implementing VA-calibration separately for each country can enhance estimation accuracy. However, the number of CHAMPS cases is quite limited in some countries. The CHAMPS dataset used here has a total of 570 cases across 6 countries for a $5 \times 5$  misclassification matrix, which leaves less than $4$ cases per cell on average. Combined across countries, 6 or fewer samples are observed for $75\%$ of the MITS-VA cause pairs, while $32\%$ are unobserved (See Figure~S1 in the supplement). This is grossly inadequate for unstructured estimation of $C^2$ misclassification rates in each country separately. 
    Such estimates will be highly imprecise, rendering them practically futile for any subsequent calibration of VA-only data.
\end{itemize}

\section{Method}\label{sec: Method}

In this section, we propose a comprehensive framework to mitigate the above issues. First, we propose a novel parsimonious model for estimating a homogeneous misclassification matrix for a classifier in a very low sample size setting. We then generalize to a nested modeling framework of progressively richer parameterized models by relaxing the parsimony and incorporating country-specific heterogeneity through adaptive shrinkage. 

\subsection{Base Model Based On \textit{Intrinsic Accuracy} And \textit{Systematic Preference}}%\paragraph{Exact decomposition of false positive into sensitivity and \textit{relative false positive}.} 

Let $\phimat=(\phi_{ij})$ denote the $C \times C$  misclassification matrix of COD predictions from a VA classifier (V) with respect to the MITS COD diagnosis (M). The diagonal $\phi_{ii}=\bbP \left( V=i \con M=i \right)$ is the sensitivity for cause $i$, and for $i \neq j$, the off-diagonal $\phi_{ij}=\bbP \left( V=j \con M=i \right)$ is the false positive (FP). So $\phimat$ is a Markov matrix (i.e., non-negative entries and rows add up to one). Due to the limited availability of MITS cases and the prevalence of unrecorded MITS-VA cause pairs, separately estimating the $C^2$  misclassification rates of $\phimat$ in an unstructured manner is undesirable, %would be prohibitive 
especially when extending to country-specific estimates. We propose a novel and parsimonious base model recognizing two fundamental processes for a classifier -- %intrinsic accuracy and systematic preference. we recognize two fundamental processes for modeling a homogeneous  misclassification matrix with a lower complexity:
algorithm's design in correctly identifying a MITS cause, and the systematic preference that inclines it towards favoring certain causes. 

\medskip
\noindent {\bf \textit{Intrinsic Accuracy (Diagonal Effect).}} %Since systematic preference acts independently to MITS, it can produce correct matches in addition to producing false positives. Besides,
Given VA algorithms are designed to produce correct matches, let $a_i$ denote its {\em intrinsic accuracy} for cause $i$, indicating the probability that the algorithm correctly identifies cause $i$ by design. Conceptually, intrinsic accuracy from predicting cause $i$ originates through a specific combination of symptoms reported in the VA record, which manifest only when the true cause is $i$.
% Given VA algorithms are designed to produce correct matches, let $a_i$ denote its {\em intrinsic accuracy} for cause $i$, i.e., the probability the algorithm correctly diagnoses cause $i$ by design. Broadly, intrinsic accuracy can be thought of as arising from the VA algorithm predicting cause $i$ based on a specific combination of symptoms reported in the VA record which arise only when the true cause is $i$. 
% Each cause will thus have a different intrinsic accuracy and we can stack these up in a vector $\bs{a} = (a_1,\dots,a_C)^\T$. % as the \textit{intrinsic accuracies}, the amount by which the algorithm is designed to produce correct matches. 

\medskip
\noindent {\bf \textit{Systematic Preference or Pull (Column Effect).}}
In scenarios where an algorithm fails to produce a correct match by design, our base model assumes a simplistic scenario where it assigns any of the $C$ causes with prespecified probabilities regardless of what the true (MITS) cause is. 
%incurs a false positive in favor of cause $j$ instead of $i$, 
This can be interpreted as the algorithm's \textit{systematic preference or pull} towards predicting a cause. 
For cause $j$ we denote the pull by  %In the most complex scenario, the preference differs across MITS causes. But in the simplest scenario, there is a common preference for all MITS causes. 
%let $B$ be a random variable denoting the cause influenced by the preference. %When $B$ is independent of MITS, we define $B$ as the \textit{systematic preference} and 
$\alpha_j$, and it is the probability of assigning cause $j$ in failing to match the correct cause $i$ by design.
% = \bbP(B=j)$. We refer to $\bs{\alpha} = (\alpha_1,\dots,\alpha_C)^\T$ 
%be the \textit{pull} and it is the amount by which the VA prediction is pulled systematically toward a cause $j$. 

\medskip
Intrinsic accuracy and pull encapsulate key algorithmic characteristics and govern the misclassification mechanism. Although we do not witness the mechanisms directly, the observed misclassification proportions are a combined influence of these fundamental events. %The two mechanisms of intrinsic accuracy and systematic preference 
%completely specify the  misclassification rate matrix $\phimat$. 
The intrinsic accuracies $a_i$ are {\em diagonal effects} and they model correct matches by design, adding to the diagonal entries of $\phimat$. The pulls $\alpha_j$'s, on the other hand, are {\em column effects} as they specify probabilities of adding to the columns of $\phimat$ regardless of the row (MITS cause).

\begin{figure}[!t]
     \centering
     \begin{subfigure}[b]{0.244\textwidth}
         \centering
         \includegraphics[width=\textwidth]{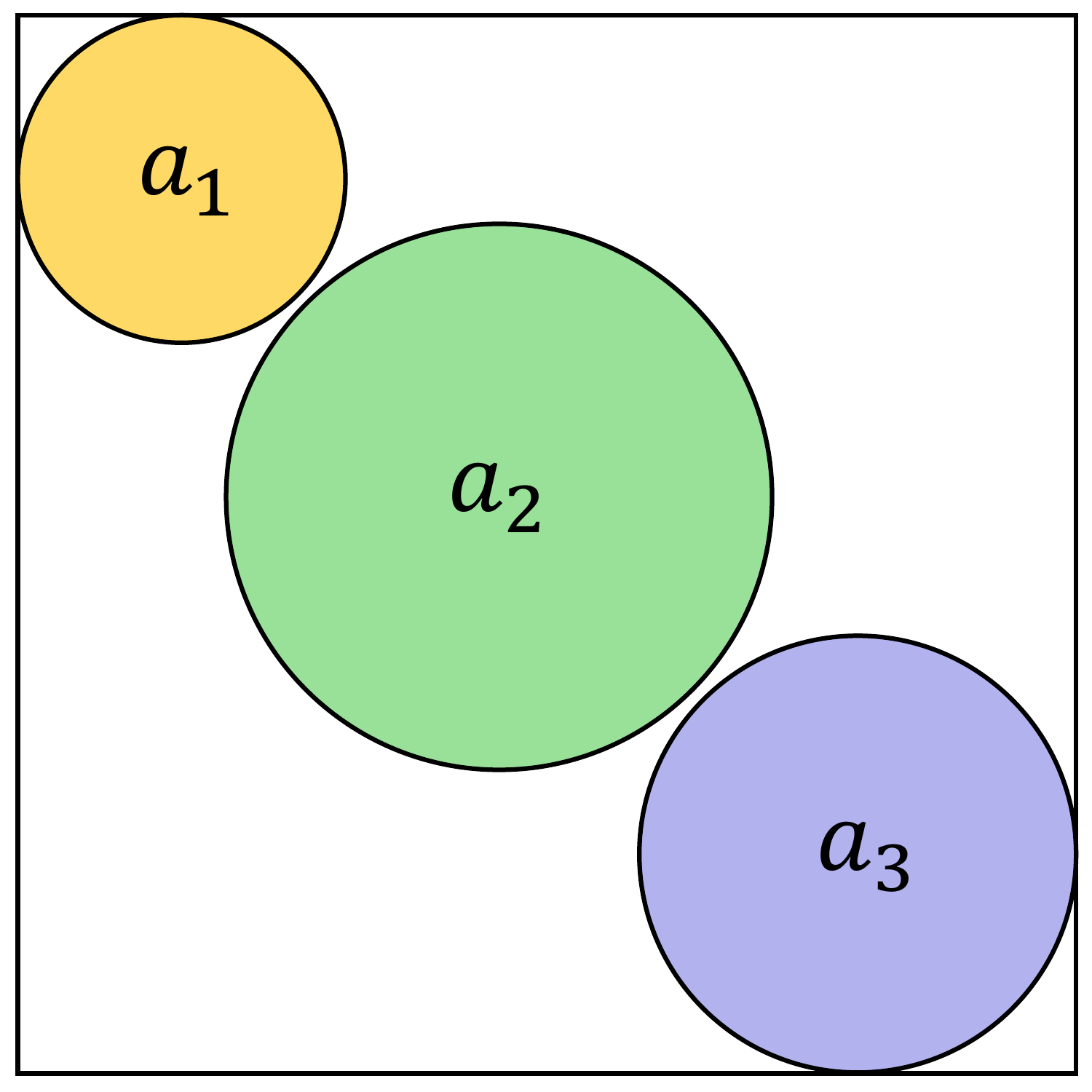}
         \caption{}
     \end{subfigure}
     \hfill
     \begin{subfigure}[b]{0.244\textwidth}
         \centering
         \includegraphics[width=\textwidth]{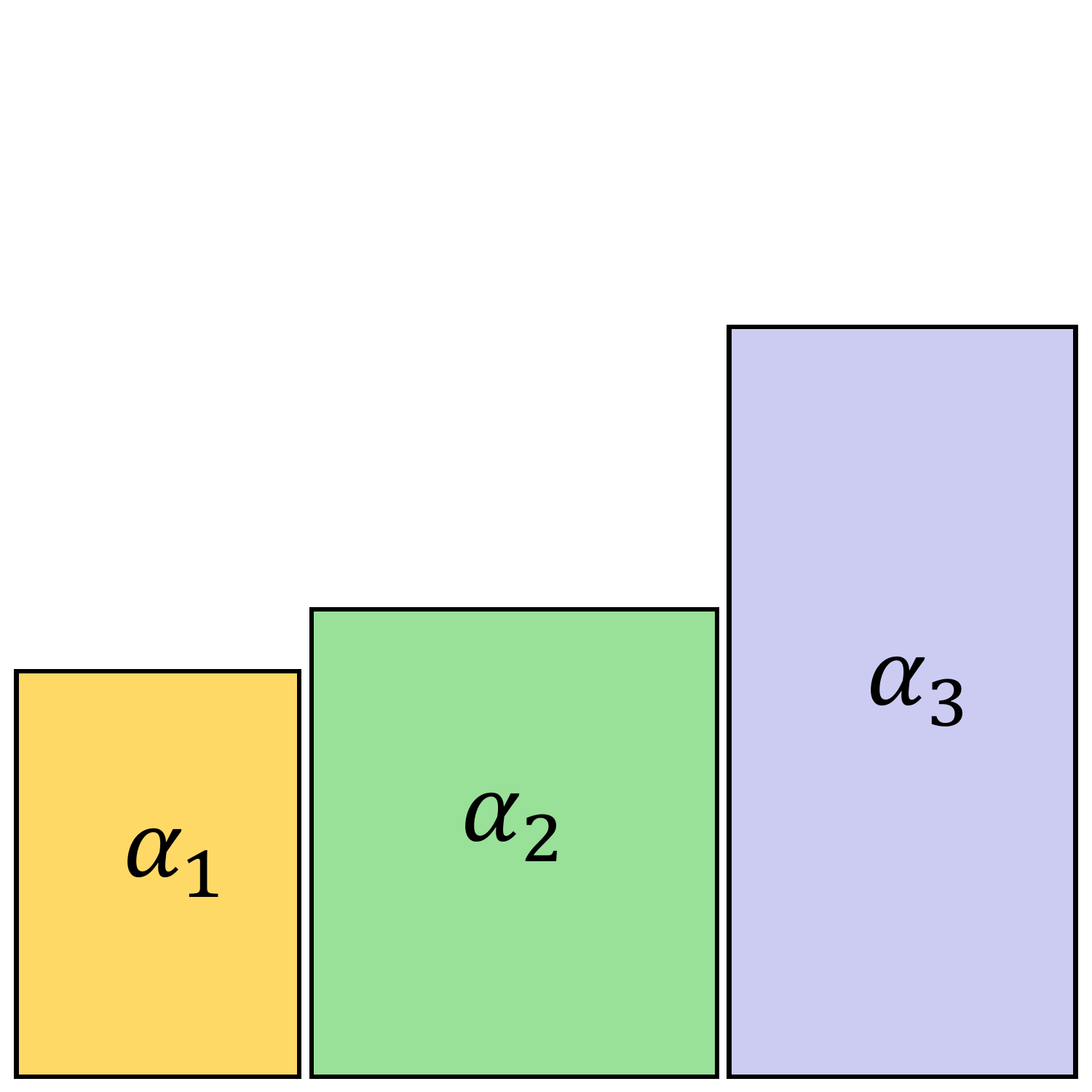}
         \caption{}
    \end{subfigure}
     \hfill
     \begin{subfigure}[b]{0.244\textwidth}
         \centering
         \includegraphics[width=\textwidth]{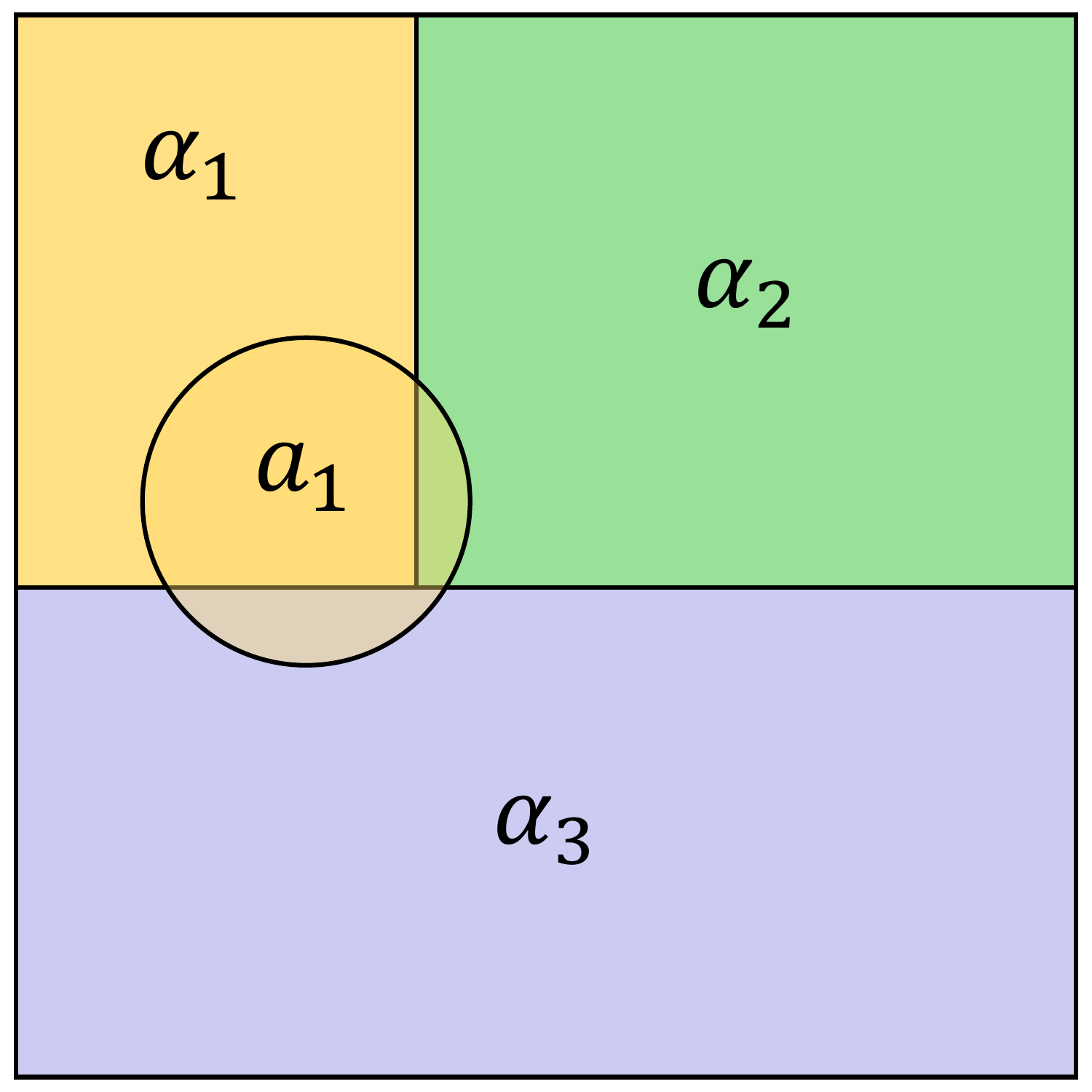}
         \caption{}
    \end{subfigure}
     \hfill
     \begin{subfigure}[b]{0.244\textwidth}
         \centering
         \includegraphics[width=\textwidth]{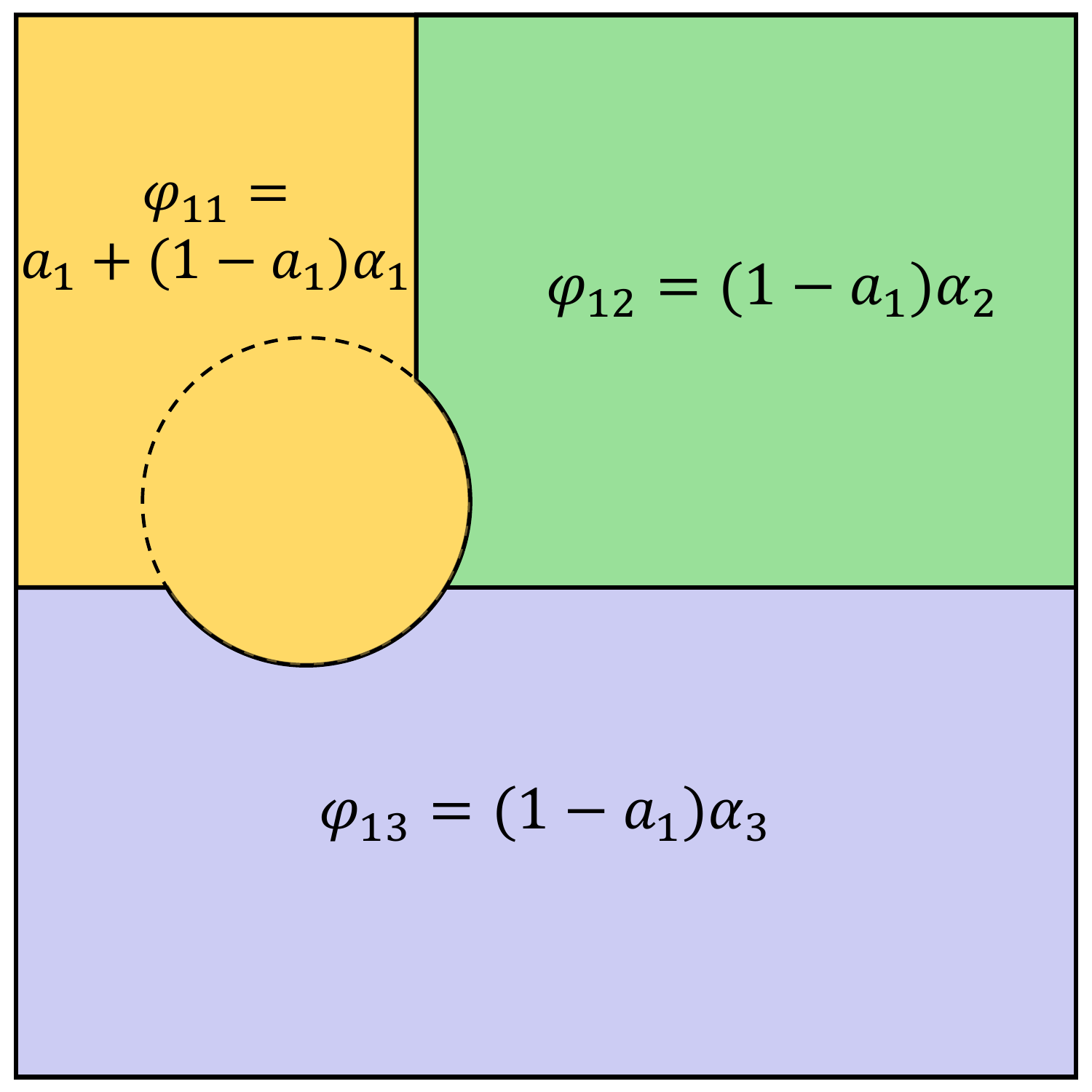}
         \caption{}
    \end{subfigure}
    \caption{For 3 causes, (a) and (b) show intrinsic accuracy and pull as diagonal and column effects. The Venn diagrams in (c) and (d) collectively illustrate the mechanism that defines sensitivity and false positives in the base model for MITS cause 1. A similar mechanism applies to other MITS causes. Causes are coded in different colors.}\label{fig: venn diagram}
\end{figure}

The mechanism is formally depicted in Figure \ref{fig: venn diagram}. A correct match for true cause $i$ occurs either by design (with probability $a_i$), or by the pull $\alpha_i$ towards cause $i$ in failing to match by design (with probability $(1-a_i)\alpha_i$). For the same true cause, a misclassification into cause $j$ occurs only by the pull $\alpha_j$ towards cause $j$ after it fails to match by design (with probability $(1-a_i)\alpha_j$). So for $a_i \in [0,1]$, $\alpha_j \in [0,1]$, and $\sum_{j=1}^C \alpha_j =1$ we have:
\begin{equation}\label{eq:base}
    \begin{split}
        \phi_{ii} &= a_i + (1-a_i)\alpha_i \quad  \forall i,\\
        \phi_{ij} &= (1-a_i)\alpha_j \quad \forall i,j \neq i.
    \end{split}
\end{equation}
% \begin{align}\label{eq:base}
%     \phi_{ii} &= a_i + (1-a_i)\alpha_i \quad  \forall i, \nonumber \\
%     \phi_{ij} &= (1-a_i)\alpha_j \quad \forall i,j \neq i, \text{ where}\\ 
%     a_i & \in [0,1], \text{ } \alpha_j \in [0, 1], \text{ and } \sum_{j=1}^C \alpha_j =1. \nonumber
% \end{align}
% \begin{align}\label{eq:base}
%     \phi_{ii} &= a_i + (1-a_i)\alpha_i \,  \forall i, \nonumber \\
%     \phi_{ij} &= (1-a_i)\alpha_j\, \forall i,j \neq i,\\ 
%     0 \leq & a_i \leq 1,\, 0 \leq \alpha_j \leq 1, \sum_{j=1}^C \alpha_j =1. \nonumber
% \end{align}
The base model offers numerous benefits in efficiently modeling the misclassification matrix. It posits a parsimonious characterization of misclassifications, requiring estimation of only $2C-1$ parameters ($C$ many $a_i$'s and $C-1$ many $\alpha_j$'s). This accrues considerable dimension-reduction contrasted with the unstructured model which has $C^2-C$ free parameters. This parsimony will later allow extension to modeling country-specific variations using very limited sample sizes. Additionally, introduction of the systematic preference enables us to explicitly measure the error that the algorithm systematically incurs (See Figure~\ref{subfig: single-neonate-insilicova-pull}). Beyond calibrating for the implied misclassification rates, %Properly addressing these errors only enhances the accuracy of our analysis. 
it also allows us to identify causes that a VA algorithm systematically favors, providing deeper insights into its functioning %. By accounting for the pull factor, we can make adjustments to compensate for systematic errors, specifically since existing algorithms like InsilicoVA are being widely implemented. This 
and opening up opportunities for designing more precise VA algorithms. %Second, as a modeling aid, pull captures commonality in false positives across MITS causes that stem from a systematic preference. This facilitates borrowing of information across MITS causes and countries, and improves the stability of estimates, particularly when dealing with a limited sample size. Finally, as discussed earlier, modeling the  misclassification matrix requires estimating $C(C-1)$ parameters when there is no systematic preference. Under strong pull, this complexity reduces to $2C-1$ parameters. This offers a substantial dimension reduction of the same order of magnitude as the number of causes, thereby significantly enhancing the stability of estimates.

Intrinsic accuracy and systematic preference are hypothesized latent mechanisms for misclassification. They cannot be directly inferred from the data as we only observe overall misclassification counts for each VA-MITS cause pair. Below we provide a theoretical justification for positing the base model. All proofs are in the supplement.%Our first theoretical result provides an equivalence between this model with a characterization of  misclassification odds that can be directly estimated from the observed  misclassification counts.

% \begin{theorem}[Characterization of base model in terms of constant odds] The  misclassification matrix $\phimat$ can be completely specified using intrinsic accuracy and systematic preference as in (\ref{eq:base}) if and only if the  misclassification odds $\phi_{ij}/\phi_{ik}$ does not depend on $i$ for all $i$ and all $j,k \neq i$.
% \end{theorem}
\begin{theorem}[Characterization of base model in terms of constant odds]\label{thm: logodds characterization} The misclassification matrix $\phimat$ can be completely specified using intrinsic accuracy and systematic preference as in (\ref{eq:base}) if and only if %for some non-negative $\theta_j$'s, 
the misclassification odds $\phi_{ij}/\phi_{ik}$ of VA predicting cause $j$ or $k$ is constant with respect to the MITS cause $i$ for any %does not depend on $i$ %equals $\theta_j/\theta_k$ for all $i$ and 
$i \neq j,k $.
\end{theorem}
%\pink{AD: Please correct the statement if I misrepresented it, and add the proof in the supplement.}
% \pink{AD: Removed additional notation of the $\theta$'s.}

\begin{figure}[!b]
    \centering
    \includegraphics[scale=0.25]{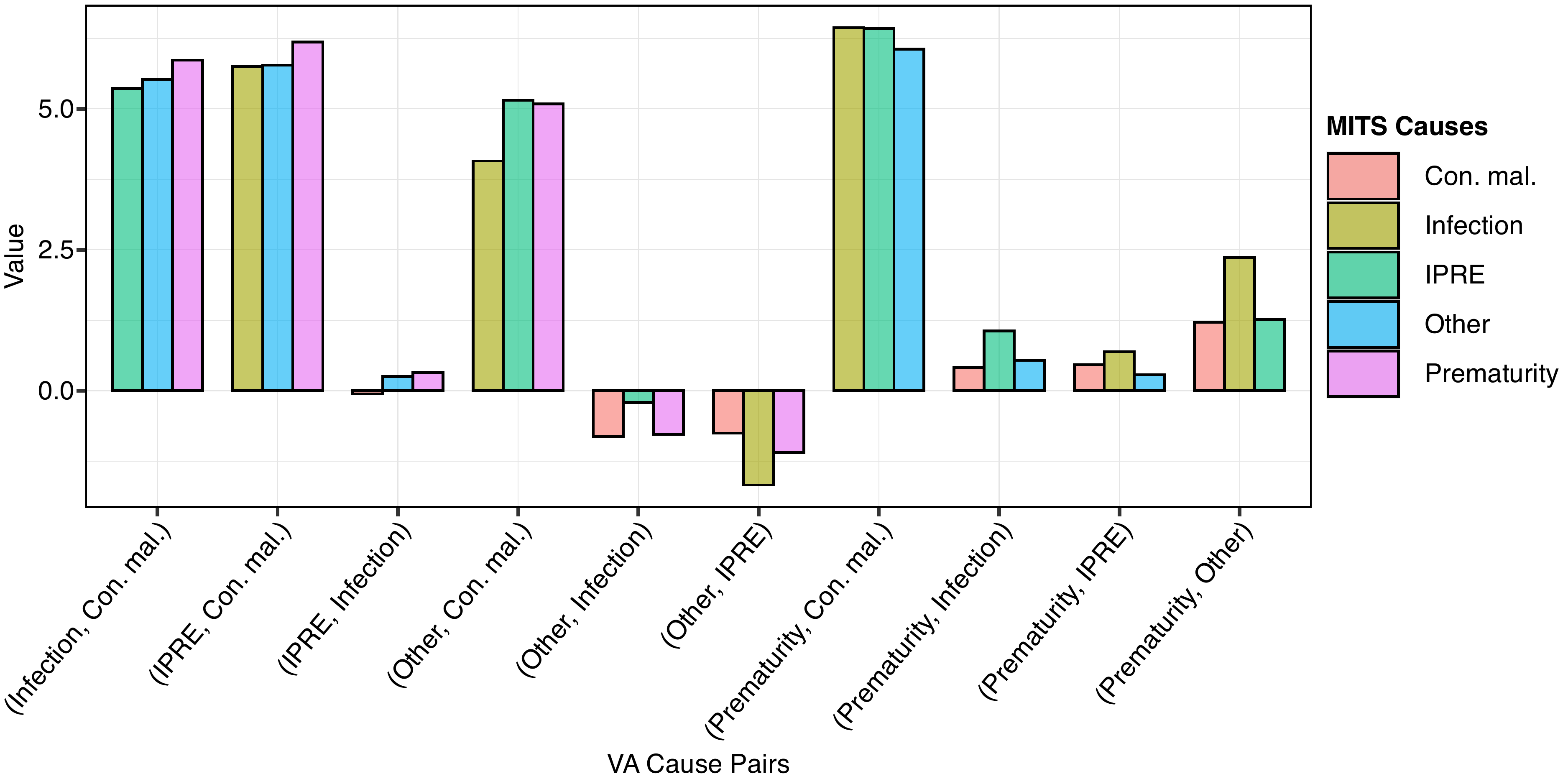}
    \caption{Similarity of  misclassification log odds $\log \left( \phi_{ij}/\phi_{ik} \right)$ between MITS causes for all VA cause pairs. The horizontal axis shows VA cause pairs  $(j,k)$. For each cause pair, the grouped bar plot presents log odds for all MITS causes $i \neq j,k$. Con. mal. denotes Congenital malformation. %\pink{AD: Provide caption and update the figure as needed.}
    }
    \label{fig: logodds}
\end{figure}
The Theorem states that the base model (\ref{eq:base}) can be equivalently characterized as the odds of the VA algorithm misclassifying an MITS cause into cause $j$ or $k$, where the odds are only a function of $(j,k)$ and do not depend on the MITS cause. This provides an equivalence between the base model and misclassification odds that can be directly estimated from the observed misclassification counts. Figure~\ref{fig: logodds} plots the log-odds for the neonatal CHAMPS data from InSilicoVA. We see a strong pattern --- for each pair $(j,k)$ of VA causes the odds $\phi_{ij}/\phi_{ik}$ are relatively constant across different MITS causes $i$. Figure~\ref{fig: logodds} combined with the characterization in Theorem~\ref{thm: logodds characterization} suggests strong evidence in favor of the base model.  %As Theorem 1 establishes the equivalence of this phenomenon with the hypothesized mechanisms of intrinsic accuracy and systematic preference, this provides strong evidence in favor of the base model. 

The next result proves that given only observed data on the misclassification counts, the parameters specifying intrinsic accuracy and systematic preference can be identified. 
Let $\{(V_r,M_r)\}_{r \in s}$ denote the paired VA- and MITS-COD for CHAMPS cases in site (country) $s$. Let us define the observed  misclassification count matrix in country $s$ as
\begin{equation}\label{eq: tmat}
    \tmat_s = \left(  t_{sij} \right) = \sum_{r \in s} \, \bbI \left( V_r = j \con M_r = i \right),
\end{equation}
and the pooled misclassification matrix $\tmat = \sum_{s=1}^S \tmat_s$. Let $n_{si} = \sum_{r \in s} \mathbb I(M_r =i)$ denote the number of cases in country $s$ with MITS cause $i$ and $n_i = \sum_{s=1}^S n_{si}$ is the total sample size for MITS cause $i$ across all CHAMPS sites. Let  $\tmat_{i*} = \left(t_{i1},\dots,t_{iC}\right)^\T$ and $\phimat_{i*} = \left(\phi_{i1},\dots,\phi_{iC}\right)^\T$ denote the $i^{th}$ row of $\tmat$ and $\phimat$. Then we can model the observed  misclassification counts as 
    \begin{equation}\label{eq:counts}
        \tmat_{i*} \overset{ind}{\sim} \multinomial(n_i, \phimat_{i*}), \quad \forall i = 1, \dots, C.
    \end{equation}%Assuming observed classifications are conditionally independent across MITS causes, for all $i = 1, \dots, C$ we hierarchically specify
%\begin{align}
%    \tmat_{i*} & \overset{ind}{\sim} \multinomial(\phimat_{i*}),\, \text{with} \,\, \phi_{ii} = 1 - (1 - a_i)(1 - \alpha_i) \,\, \text{and} \,\, \phi_{ij} = (1 - a_i)\alpha_j \,\, \forall \, i \neq j, \label{eq: homogeneous model strong pull}\\
%    a_i &\overset{iid}{\sim} \betaprior \left(b,d\right) \text{  with  } b,d>0,  \label{eq: prior intrinsic accuracy}\\
%    \bs{\alpha} &\sim \dirichlet \left( \bs{e} \right) \text{  with  } e_1, \dots,e_C>0.  \label{eq: prior pull}
%\end{align}
%We refer to (\ref{eq: homogeneous model strong pull})--(\ref{eq: prior pull}) as the \textit{homogeneous  misclassification modeling under a strong pull}. Here $\tmat_{i*} = \left(t_{i1},\dots,t_{iC}\right)^\T$ and $\phimat_{i*} = \left(\phi_{i1},\dots,\phi_{iC}\right)^\T$ are the $i^{th}$ row of $\tmat$ and $\phimat$, and they correspond to MITS cause $i$. In this case, we have the following two results.
\begin{proposition}[Identifiability of base model]\label{prop: identifiability strong pull}
Under the base model (\ref{eq:base}) for the misclassification rate matrix, the likelihood (\ref{eq:counts}) is identifiable with respect to the $a_i$'s and $\alpha_j$'s.
\end{proposition}
The proposition confirms that under the base model, we can uniquely recover intrinsic accuracies and pull that govern the misclassification mechanism. The result validates the base model and implies that the accuracy and pull estimates will be asymptotically consistent. %The identifiability result validates the base model and implies that the accuracy and pull estimates will be asymptotically consistent. 
We do not pursue a formal proof of consistency as the sample size is very low. %We do not pursue a formal proof of consistency as we have a very small sample scenario. %This validates the base model. 

% \begin{proposition}[Convexity of likelihood under strong pull]\label{prop: convexity strong pull}
%     {\color{blue} to be included} 
% \end{proposition}
% \pink{AD: it's high time we either add the proof of this result or remove it altogether (it is not too important).}

% Proposition~\ref{prop: identifiability strong pull} confirms that under strong pull the model can uniquely recover intrinsic accuracies and pull that govern the classification mechanism. Proposition~\ref{prop: convexity strong pull} proves that there exists a unique intrinsic accuracies $\bs{a}$ and pull $\alpha$ that maximizes the likelihood. This makes the optimization computationally tractable. Together, the propositions prove the validity and efficacy of the model. %Proposition \ref{prop: identifiability strong pull} provides the assurance that the base model parameters can be estimated relaibly from observed  misclassification counts. 

\subsection{A General Homogeneous Model}\label{sec:pooled}

Despite the evidence in Figure (\ref{fig: logodds}) for InsilicoVA's misclassification among neonatal deaths in CHAMPS, it might be too restrictive in practice to assume that misclassification rates exactly satisfy the base model (\ref{eq:base}).
% In practice, the assumption of the misclassification rates being exactly structured as in the base model (\ref{eq:base}) might be too strong, despite the evidence in Figure (\ref{fig: logodds}) that it can indeed be an excellent choice for the observed VA  misclassification rates. 
Also, the extent to which misclassification rates satisfy this structure may vary across age groups and choice of VA algorithms. Instead, we propose a more general model that shrinks towards the base model to achieve parsimony.

First, we decompose the false positive rates as
\begin{equation}\label{eq: sens rfp decomposition}
    \phi_{ij}=\left(1- \phi_{ii}\right) q_{ij} \quad \forall i \neq j,
\end{equation}
where $\phi_{ii}$ is the sensitivity and $q_{ij}=\bbP \left( V=j \con M=i, V \neq i \right)$ is defined as the \textit{relative false positive (relative FP)}. Note that, under base model (\ref{eq:base}), we have \begin{equation}\label{eq: systematic rfp}
q_{ij} = \frac{\bbP \left( V=j, M=i \right)}{\sum_{k \neq i} \bbP \left( V=k, M=i \right)} =  \frac{(1-a_{i})\alpha_j}{(1-a_{i})\sum_{k \neq i} \alpha_k} = \frac{\alpha_j}{1 - \alpha_i} := \alpha^*_{ij}.
\end{equation}
We propose probabilistic models for both sensitivities and relative FP, and shrink them towards respective expressions in the base model. The complete Bayesian hierarchical model for modeling a general homogeneous misclassification matrix is given by
\begin{align}\label{eq:pooled}
    \tmat_{i*} & \overset{ind}{\sim} \multinomial(\phimat_{i*}), \text{  with  } \phi_{ij}=(1-\phi_{ii}) q_{ij} \quad \forall i \neq j, \nonumber \\ %\label{eq: homogeneous model}\\
    \phi_{ii} &\overset{ind}{\sim} \betaprior \left( 0.5 + \kappa (a_i +(1- a_i) \alpha_i), 0.5 + \kappa (1- a_i) (1-\alpha_i)\right) \text{  with  } f,\kappa>0, \nonumber  \\ %\label{eq: prior sensitivity homogeneous}\\
    \bs{q}_{i *} &\overset{ind}{\sim} \dirichlet \left( 0.5 + \lambda \bs{\alpha}^*_{i *} \right) \text{  with  } g,\lambda>0, %\label{eq: prior rfp homogeneous}
     \\
     a_i &\overset{iid}{\sim} \betaprior \left(b,d\right) \text{  with  } b,d>0,  %\label{eq: prior intrinsic accuracy} 
     \nonumber \\
    \bs{\alpha} &\sim \dirichlet \left( \bs{e} \right) \text{  with  } e_1, \dots,e_C>0. \nonumber % \label{eq: prior pull}
\end{align}
where $\bs{\alpha}^*_{i *} = \left(\alpha^*_{i1},\dots,\alpha^*_{i,i-1},\alpha^*_{i,i+1},\dots,\alpha^*_{iC}\right)^\T$ with $\alpha^*_{ij}$ defined in (\ref{eq: systematic rfp}).

This model for the pooled data assigns a Beta prior on the sensitivities $\phi_{ii}$ and a Dirichlet prior on the relative FP $q_{ij}$ each centered around their respective base models. This ensures that the resulting prior for $\phimat$ assigns positive mass to every interior point of the class of $C \times C$ Markov matrices, and given enough data, the posterior will concentrate around the true misclassification matrix. However, in settings with limited data, the choice of a structured prior is important. 
If $\kappa \uparrow \infty$ and $\lambda \uparrow \infty$ in (\ref{eq:pooled}), the model for $\phimat$ degenerates exactly to the base model.  %simplifies to modeling under the strong pull. 
On the other extreme, as $\lambda, \kappa \downarrow 0$ the priors respectively converge to \textit{Jeffreys non-informative priors} $\betaprior(0.5,0.5)$ and $\dirichlet(0.5, \dots, 0.5)$ that are often utilized for unstructured modeling of proportions. Thus the addition of just two additional parameters, $\kappa$ and $\lambda$, introduces flexibility into the framework %and naturally accommodating the gamut a wide range of \textit{pull strengths} to account for deviation from the strong pull. Through this adaptability, the framework encompasses diverse preference patterns of an algorithm in a nested structure,
producing a wide spectrum of models with varying degrees of structure. %This also enables the model to systematically transition from a simpler configuration with systematic preference to a more intricate one with no preference if supported by the data. 

\subsection{Country-specific Misclassification Model}\label{sec:country} 

%Despite the current availability of CHAMPS data across multiple countries, the current implementation of VA-calibration aggregates data from all countries and provides a \textit{global} estimate of  misclassification matrix for all. Contrary to this, 
Figure~\ref{fig: observed classification proportions} suggests that the  misclassification rates of %sensitivities for some MITS causes and false positive rates for some MITS-
VA with respect to MITS can substantially vary across countries for some cause-pairs. However, due to the limited availability of CHAMPS data per country, it is not feasible to perform a separate unstructured estimation of misclassification rates for each country. Thus a pooled estimation of misclassification rates under a homogeneous model, though not ideal, was previously resorted to as the only viable approach. In this regard, the base model has been designed to address the challenges posed by limited sample sizes by leveraging any discernible structure in the matrix. This significantly reduces the model's dimensionality, allowing us to introduce heterogeneity and generate country-specific estimates for misclassification rates.

%this Since data is aggregated across countries, homogeneous estimates tend to exhibit higher precision (lower variance). But in the presence of strong heterogeneity the aggregation makes the estimate locally incongruous, which substantially increases their bias. We can implement homogeneous  misclassification modeling separately for each country to account for the heterogeneity, but it is undesirable for two key reasons. First, it prohibits information sharing across countries, which is counterintuitive since MITS and VA predict the same set of causes in all countries. Second, the sample size observed across countries is very limited. Although the naive separate modeling will improve country-specific estimation accuracy (reduce bias), the limited number of samples will increase their variance (reduce precision), resulting in unstable estimates. Besides, with certain MITS causes being unobserved in some countries, the separate modeling cannot provide estimates for them. Together, these underscore the importance of incorporating common patterns in the matrix and leveraging sharing of information across causes and countries.

%\paragraph{Country-specific (heterogeneous) modeling of  misclassification matrix.} 
Let $\phimat_s = (\phi_{sij})$ denote the  misclassification matrix for country $s$ and $\tmat_s$ is defined as in (\ref{eq: tmat}). %Since all MITS causes are not observed in all countries, let $\mathcal{O}_s$ denote the set of MITS causes that are observed in country $s$. 
%Assuming the labeled data are observed independently across countries, for $s = 1, \dots, S$ we hierarchically 
We propose a heterogeneous model by introducing country-specific probabilistic models for sensitivities and relative FP centered around their respective pooled models in (\ref{eq:pooled}): 
\begin{align}\label{eq:het}
    \tmat_{si*} &\overset{ind}{\sim} \multinomial(n_{si},\phimat_{si*}) \text{  with  } \phi_{sij}=(1-\phi_{sii}) q_{sij} \quad \forall i, %\in \mathcal{O}_s \text{  and  } i \neq j, 
    \nonumber \\%\label{eq: heterogeneous model}\\
    \phi_{sii} &\overset{ind}{\sim} \betaprior \left( 0.5 + \gamma \phi_{ii}, 0.5 + \gamma (1- \phi_{ii})\right) \quad \forall i \text{  with  } f,\gamma>0, \\
    \bs{q}_{si *} &\overset{ind}{\sim} \dirichlet \left( 0.5 + \delta \bs{q}_{i *} \right) \quad \forall i \text{  with  } g,\delta>0. \nonumber %\label{eq: prior rfp hetergeneous}.
\end{align}
We complete the hierarchy by setting the same priors for the pooled parameters $\phi_{ii}$, $\bs{q}_{i *}$, $a_i$ and $\bs{\alpha}$ as in (\ref{eq:pooled}). %according to (\ref{eq: prior sensitivity homogeneous})--(\ref{eq: prior rfp homogeneous}), and $a_i$ and $\bs{\alpha}$ as in (\ref{eq: prior intrinsic accuracy})--(\ref{eq: prior pull}). $\tmat_{si*}$ and $\phimat_{si*}$ correspond to MITS cause $i$ in country $s$, and respectively denote the $i^{th}$ row of $\tmat_s$ and $\phimat_s$. In this specification, $\phi_{ij}$ is interpreted as the \textit{fixed effect} of MITS-VA cause pair $(i,j)$, $\phimat$ as the fixed effect of the country-specific  misclassification matrices that is common across countries.
To account for heterogeneity across countries, $\phi_{sii}$ and $\bs{q}_{si *}$ can be interpreted as the country-specific \textit{random effects} of sensitivities and relative FP. The random effects are respectively distributed according to Beta and Dirichlet distributions and pivot around the respective \textit{fixed effects} from the homogeneous model (\ref{eq:pooled}). The deviations of random effects from fixed effects quantify the \textit{degree of heterogeneity}. 

\begin{figure}[!b] %  figure placement: here, top, bottom, or page
   \centering
   \includegraphics[width=\textwidth]{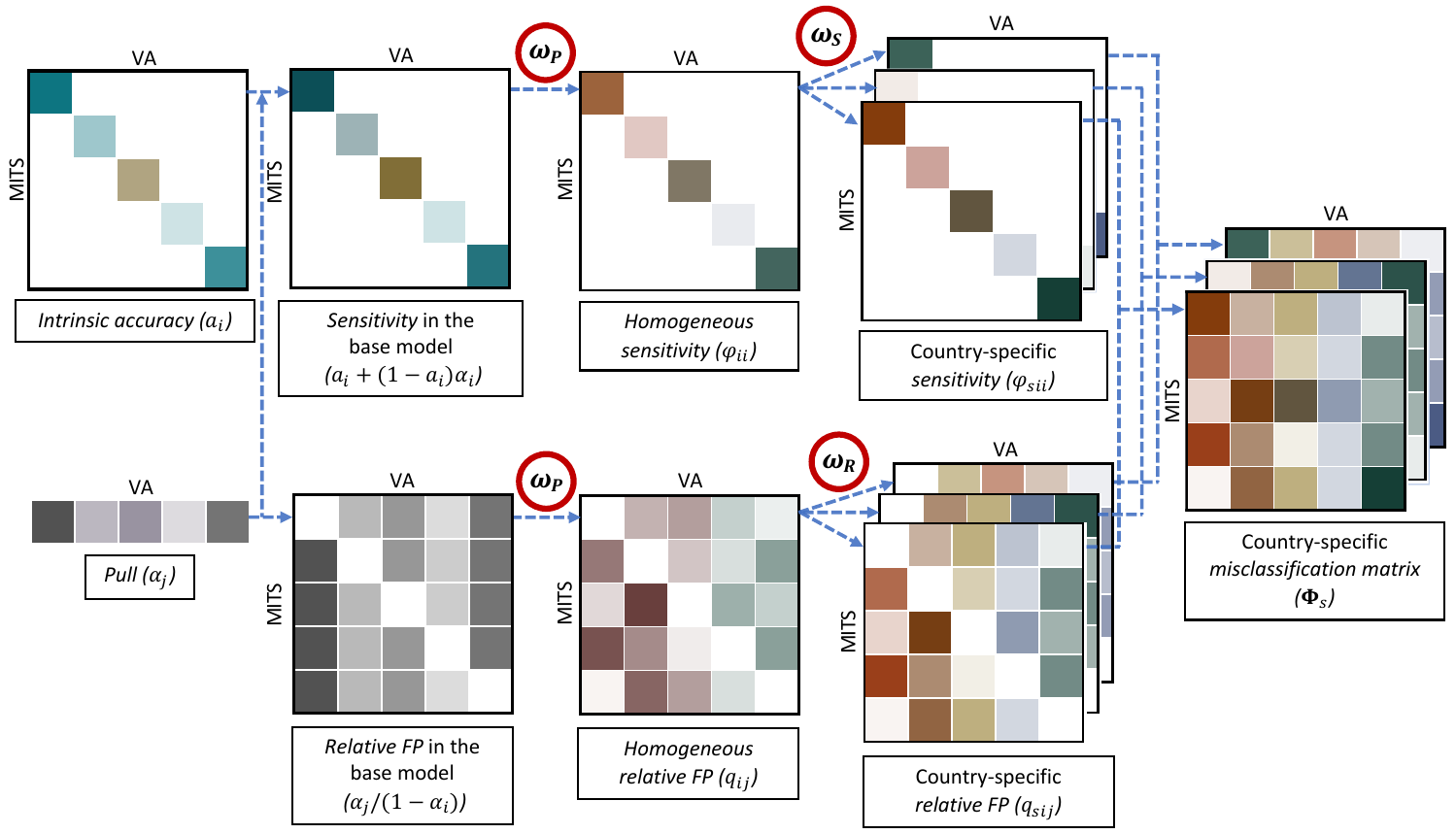}
   \caption{A schematic diagram of country-specific  misclassification modeling framework.% \pink{AD: The schematic has the parameters $\omega_p$, $\omega_r$ etc. which are not yet introduced! Also, I have removed much of the jargons that were used in the figure, please write in terms of the base model, homogenous model and heterogenous model.}
   }
   \label{fig: schematic diagram}
\end{figure}
%As the variability around them accounts for the heterogeneity effects concentrate at respective fixed effects,
The heterogeneous model (\ref{eq:het}) subsumes the homogeneous model (\ref{eq:pooled}) as a specific instance as $\gamma,\delta \to \infty$. Conversely, as $\gamma, \delta \downarrow 0$, the random effects are distributed independently as the non-informative priors $\betaprior(0.5, 0.5)$ and $\dirichlet(0.5, \dots, 0.5)$, respectively. In this case, no information is shared between countries and it is analogous to modeling misclassifications separately for each country. This framework thus encompasses separate country-specific modeling as a special case. For the intermediate scenario with some heterogeneity, the fixed effects capture the average misclassification rates common across the countries, and the heterogeneity is accounted for by the variability around them. Through $\gamma$ and $\delta$, the framework thus coherently includes varied degrees of heterogeneity in a nested structure. %, allowing it to flexibly expand from the base model to the general homogeneous model to the more intricate heterogeneous model as needed.

Figure~\ref{fig: schematic diagram} presents a schematic of the proposed nested framework of the models with progressively increasing complexity. The heterogeneous model subsumes the homogeneous model which in turn incorporates the base model as a special case. 

% \subsection{Effect Sizes for Pull and Homogeneity}\label{sec:effect}
\subsection{Interpretable Model Parameters}

All parameters in the proposed framework are interpretable. The $\phi_{ij}$'s and $\phi_{sij}$'s are homogeneous and country-specific misclassification rates. The parameters specifying intrinsic accuracy ($a_i$) and systematic preference or pull ($\alpha_j$) in the base model provide an understanding of the latent mechanism dictating the VA algorithm's misclassifications. The extensions to the homogeneous model (\ref{eq:pooled}) and the heterogenous model (\ref{eq:het}) are respectively controlled by dispersion parameters $(\kappa,\lambda)$ and  $(\gamma,\delta)$ which have the following interpretation.% which can be interpreted as follows. 

Dispersion parameters in Beta or Dirichlet priors carry the interpretation of prior sample sizes for \textit{Binomial} or \textit{Multinomial} data. 
Thus $\kappa$ and $\lambda$ can be interpreted as the \textit{total prior sample size} for the prior that the pooled misclassification rates are generated from the base model. Similarly, $\gamma$ and $\delta$ provide the prior sample size for a homogeneous prior for the heterogeneous misclassification rates. Following this we parameterize them as
\begin{equation}\label{eq: effect size parameterization}
    \kappa=2 \omega_P, \quad \lambda=(C-1) \omega_P, \quad \gamma=2 \omega_S \quad \text{and} \quad \delta=(C-1) \omega_R,
\end{equation}
where $\omega_P$, $\omega_S$, and $\omega_R$ are prior sample sizes for each category (2 categories for sensitivity representing correct or incorrect match, and $C-1$ categories for relative FP representing false positive causes). Thus $\omega_P$, $\omega_S$, and $\omega_R$ are effect sizes of the framework and they respectively control the \textit{pull strength} and the degrees of homogeneity in sensitivity and relative FP. A large $\omega_P$ indicates a {\em strong pull} -- strong shrinkage towards the base model with systematic preference or pull, whereas a small value suggests {\em no pull} -- an absence of structure in the misclassification rates. %We can choose different prior sample sizes per category for sensitivity and relative FP in $\kappa$ and $\lambda$, but here we assume they are the same. In the absence of further knowledge and given the limited availability of samples, this inherently assumes \textit{symmetric deviation} from the strong pull. Similarly, large and small values of $\omega_S$ and $\omega_R$ respectively indicate homogeneity and heterogeneity in sensitivity and relative FP. 
A large value of $\omega_S$ suggests homogeneity, favoring a strong shrinkage of the country-specific sensitivities $\phi_{sii}$ towards the pooled sensitivities $\phi_{ii}$. A similar interpretation holds for $\omega_R$ for the country-specific relative FP $q_{sij}$.

\subsection{Shrinkage Priors on Effect Sizes for Pull and Homogeneity}\label{sec:prior}

Considering the significance of dispersion parameters $\omega_P$, $\omega_S$, and $\omega_R$ in regulating the pull strength and degree of homogeneity, we propose to learn them from the data. In a Bayesian approach, this necessitates assigning priors for each of them.
%\subsection{Shrinkage priors}\label{sec:prior}  %To counteract limited number of MITS cases as depicted in Figures \ref{fig: observed classification proportions} and S1 in the supplement, 
Specifically, we utilize shrinkage priors to favor a model with lower complexity, allowing more precise %\sandy{precise or stable?} 
estimation even in low-sample size settings. We transform the effect sizes to $[0,1]$ using the mapping $f(x)=1/(1+x)$ and assume that the transformed effect sizes are independently distributed as
%\begin{equation}\label{eq: prior effect size}
    $\betaprior(\varepsilon, \varepsilon), \text{ for  } \varepsilon \in (0,1)$.
%\end{equation}
On the transformed scale, 0 signifies the lowest complexity (strong pull and homogeneity) while the highest complexity corresponds to 1 (no pull and heterogeneity). With modes at 0 and 1, the prior allocates more probability near the two interpretable extremes. This encourages the model to shrink towards simplicity in low-sample size settings while being adaptable to a higher complexity if supported by data.
% \begin{figure}[!t]
%      \centering
%      \begin{subfigure}[b]{0.45\textwidth}
%          \centering
%          \includegraphics[width=.9\textwidth]{figures/prior-effectsize-transformed.pdf}
%          \caption{Transformed scale}
%      \end{subfigure}
%      \hfill
%      \begin{subfigure}[b]{0.45\textwidth}
%          \centering
%          \includegraphics[width=.9\textwidth]{figures/prior-effectsize.pdf}
%          \caption{Original scale}
%      \end{subfigure}
%     \caption{(a) on the left depicts the prior on transformed effect sizes for $\varepsilon=0.5$. For the same choice, Figure (b) on the right presents the implied prior on the original scale. %\pink{AD: Use the same style of density plots for all prior figures (the latter plots shades the regions with blue.)}
%         }
%         \label{fig: prior effect size}
% \end{figure}

Figure~S3 in the supplement shows an example of the prior for $\varepsilon=0.5$. On the transformed scale, it assigns 50\% probability outside (0.15, 0.85) and distributes the rest of the probability roughly evenly within the interval. On the original scale, this is equivalent to having at least 5.5 prior samples for each cause or category to indicate a strong pull or homogeneity, and fewer than 0.15 observations for suggesting no pull or heterogeneity.

\begin{figure}[!t]
   \centering
   \includegraphics[width=\textwidth]{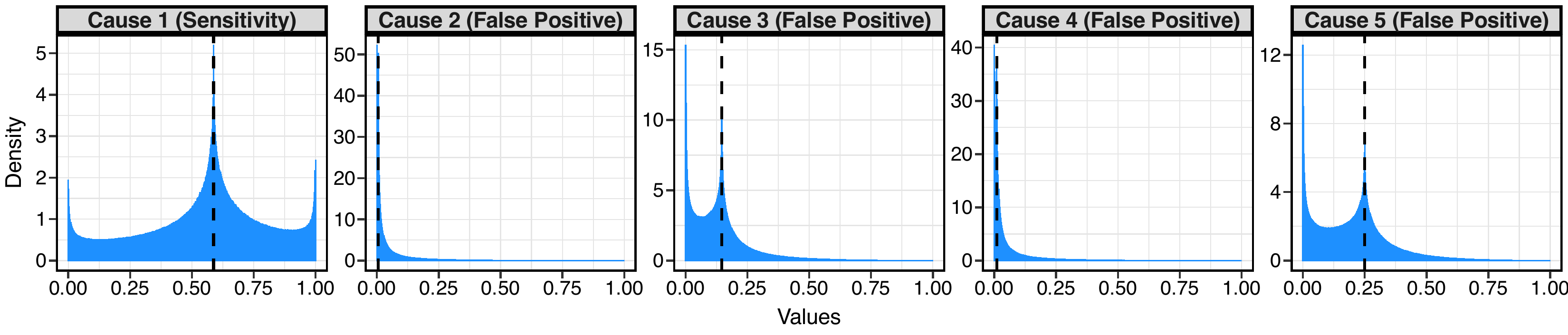}
   \caption{Conditional prior on misclassification rates in the homogeneous model (\ref{eq:pooled}) given the base model parameters and the % 5 causes with the true Cause 1 and the shrinkage prior with $\varepsilon=0.5$ assumed
   $\betaprior(0.5,0.5)$ prior on pull strength ($\omega_P$). Vertical dashed lines indicate the respective values in the base model (\ref{eq:base}).}\label{fig: conditional prior sensitivity&fp given pull} %, the figure shows the prior implied on sensitivity (Panel 1) and false positive rates (Panel 2--5) in the general homogeneous model given intrinsic accuracy and pull. Vertical dashed lines indicate the sensitivity and false positives in the base model (\ref{eq:base}). The same prior is assumed on the country-specific rates when the same shrinkage prior is assumed on the degrees of heterogeneity $\omega_S$ and $\omega_R$. In that case, the vertical dashed lines represent sensitivity and false positives in the general homogeneous model.}\label{fig: conditional prior sensitivity&fp given pull}
\end{figure}

This choice of shrinkage prior on dispersion parameters implies a conditional shrinkage on misclassification rates. Figure~\ref{fig: conditional prior sensitivity&fp given pull} shows an example of the conditional prior implied on homogeneous misclassification rates $\phi_{ij}$ in the homogeneous model (\ref{eq:pooled}) given the base model parameters $\bs{a}$ and $\bs{\alpha}$ and the $\betaprior(0.5,0.5)$ prior on $\omega_P$. For sensitivities, the conditional prior is trimodal, with peaks at the base model, 0 (shrinkage to perfect misclassification) and 1 (shrinkage to perfect classification). For false positives, the conditional prior is bimodal shrinking either to the base model or to 0. The prior thus encourages a \textit{continuous shrinkage} \citep{CARVALHO2010,Park2008}: concentrating around the base model (dashed vertical lines), perfect misclassification, or perfect classification (for sensitivities only). %and the nearest interval endpoints, and decreases as we move away from these values. These behaviors are desirable, shrinking either to base model, complete misclassification, or towards perfect classification in absence of adequate data. 
The same conditional prior is also assumed in (\ref{eq:het}) for country-specific misclassification rates $\phi_{sij}$ given their homogeneous values $\phi_{ij}$.
% Similarly, the conditional prior on the country-specific misclassification rates $\phi_{sij}$ given the homogeneous rates $\phi_{ij}$ (fixed effects) as in (\ref{eq:het}) will have similar tri- and bi-modal structure. 

When dealing with limited samples, it conveniently encourages a continuous adjustment in pull strength and degree of heterogeneity to favor a simpler model for misclassification rates, while allowing for evidence accumulation in support of a more complex model as the sample size increases. This dynamically adjusts the modeling complexity, providing an optimal balance to the bias-variance trade-off in country-specific estimation by utilizing both local and global misclassification patterns in a data-driven manner. 
%This improves the \textit{validity} (accuracy and precision) of the resulting misclassification rate estimates. %The shrinkage adds valuable adaptive learning capabilities to the framework, allowing it to detect effect sizes while capturing shared characteristics among the MITS causes. 

% \begin{figure}[h]
%    \centering
%    \includegraphics[width=.9\textwidth]{figures/prior-sensitivity&fp-hom-marginal.pdf}
%    \caption{{\bf Marginal prior on country-specific sensitivities and false positives.} For 5 causes and default recommendation of prior parameters, this is an example of prior on sensitivity (left panel) and false positive (right panel) in each country.}\label{fig: marginal prior sensitivity&fp given pull}
% \end{figure}

%For a default implementation of the proposed framework, we suggest using the following parameter values: $f=g=0.5$, which correspond to Jeffreys priors on sensitivities and relative FP rates under no pull, $b=d=1$ indicating the Uniform prior on intrinsic accuracies, $\bs{e} = \bs{1}$ assuming the Uniform prior on pull, and $\varepsilon=0.5$ for effect size priors. Practitioners are advised to analyze the impact of prior parameter selections on their specific applications and customize them accordingly. Despite the desired continuous shrinkage, the implied marginal priors on classification rates for default choices are weakly informative, closely resembling the Jeffreys prior commonly recommended in the literature as a noninformative prior choice on Binomial and Multinomial probabilities (Please see Figure S3 in the supplement).

\subsection{Extrapolation to Unobserved MITS Causes and Countries}\label{sec:pred}

On account of borrowing information across countries, our Bayesian hierarchical framework naturally allows VA misclassification rate predictions for countries not represented in the CHAMPS data. It also enables predicting misclassification rates for a CHAMPS country with unobserved MITS causes. This is quite common since CHAMPS is based on convenient sampling and certain countries frequently lack observations for some MITS causes. %, with no available labeled data for countries outside the network. 
For either task, we can obtain predictive distributions of the misclassification rates as follows.

%The proposed framework by virtue of information sharing and hierarchical modeling readily offers uncertainty-quantified \textit{posterior predictions} in these situations. If MITS cause $i$ is never observed, the posterior predictive distributions of its 
For the homogeneous model (\ref{eq:pooled}), misclassification rates for a new country are the same as the posterior distribution of pooled misclassification rates $\phi_{ij}$. %So posterior % for sensitivity and relative FP are given by the pooled estimates. 
%\begin{equation}\label{eq: pooled predict}
%    \begin{split}
%        \phi_{ii} & \sim \betaprior \left(f+ 2 \omega_P (a_i +(1- a_i) \alpha_i),f+ 2 \omega_P (a_i +(1- a_i) \alpha_i)\right), \text{  and  }\\
%    \bs{q}_{i *} & \sim \dirichlet \left( g + (C-1) \omega_P \bs{\alpha}^*_{i *} \right).
%    \end{split}
%\end{equation}
%Here $\omega_P$, $a_i$ and $\bs{\alpha}$ are substituted by their posteriors, and $f$ and $g$ are prefixed. The predictive distributions are centered at their respective values from the base model, and its uncertainties are informed by the posterior of pull strength $\omega_P$. Finally, posterior predictive false positive rates are obtained by combining them according to the decomposition (\ref{eq: sens rfp decomposition}).
%Likewise, when MITS cause $i$ is not observed in country $s$, the framework relies on the general homogeneous model framework. In this case
For the heterogeneous model (\ref{eq:het}), the posterior predictive distribution of sensitivity and relative FP in a new country $s$ are given by draws
\begin{equation}\label{eq: het predict}
    \begin{split}
    \phi^{(r)}_{sii} & \sim \betaprior \left( 0.5 + 2 \omega_S^{(r)} \phi_{ii}^{(r)}, 0.5 + 2 \omega_S^{(r)} (1- \phi_{ii}^{(r)})\right), \text{  and  }\\
    \bs{q}^{(r)}_{si *} & \sim \dirichlet \left( 0.5 + (C-1) \omega_R^{(r)} \bs{q}^{(r)}_{i *} \right),
    \end{split}
\end{equation}
where $r$ denotes the $r^{th}$ MCMC sample. 
%Thus the posterior predictive distribution can be obtained using draws from (\ref{eq: het predict}) for every draw of the model parameters in the Markov Chain Monte Carlo run. 
%Similarly, $\omega_S$ and $\omega_R$ are substituted by their posteriors. 
Note that, the predictive distributions for a country are centered at their homogeneous values, and the uncertainties around them for sensitivity and relative FP are guided by the posterior information on degrees of heterogeneity $\omega_S$ and $\omega_R$. So the predictions from homogeneous and heterogeneous models will have similar point estimates, but the latter will have higher uncertainty that accounts for heterogeneity.

\section{Simulation Study}\label{sec: Simulation Study}

We conducted extensive simulation studies to assess the performance in the presence and absence of heterogeneity across countries. We considered three scenarios of a true misclassification matrix: $(i)$ \textit{homogeneous:} The same misclassification matrix for all countries as in (\ref{eq:pooled}), %This corresponds to (\ref{eq:pooled}) and is referred to as \textit{homogeneous}. 
$(ii)$ \textit{partly-heterogeneous:} heterogeneity only in sensitivities achieved by setting $\delta$ or $\omega_R=\infty$ in (\ref{eq:het}), % It is referred to as \textit{partly-heterogeneous}. 
and, $(iii)$ \textit{fully-heterogeneous:} heterogeneity in both sensitivity and relative FP as in (\ref{eq:het}). %We refer to this as \textit{fully-heterogeneous}. 
In each scenario, we fit these three methods that are special cases of the proposed framework. See sections S3 and S4 in the Supplement for additional details. %Details of data generation and model fitting are described in sections \ref{ssec: Default Prior Parameter Choices} and \ref{ssec: Simulation Study} in the Supplement. 

\begin{figure}[!h]
   \centering
   \includegraphics[width=\textwidth]{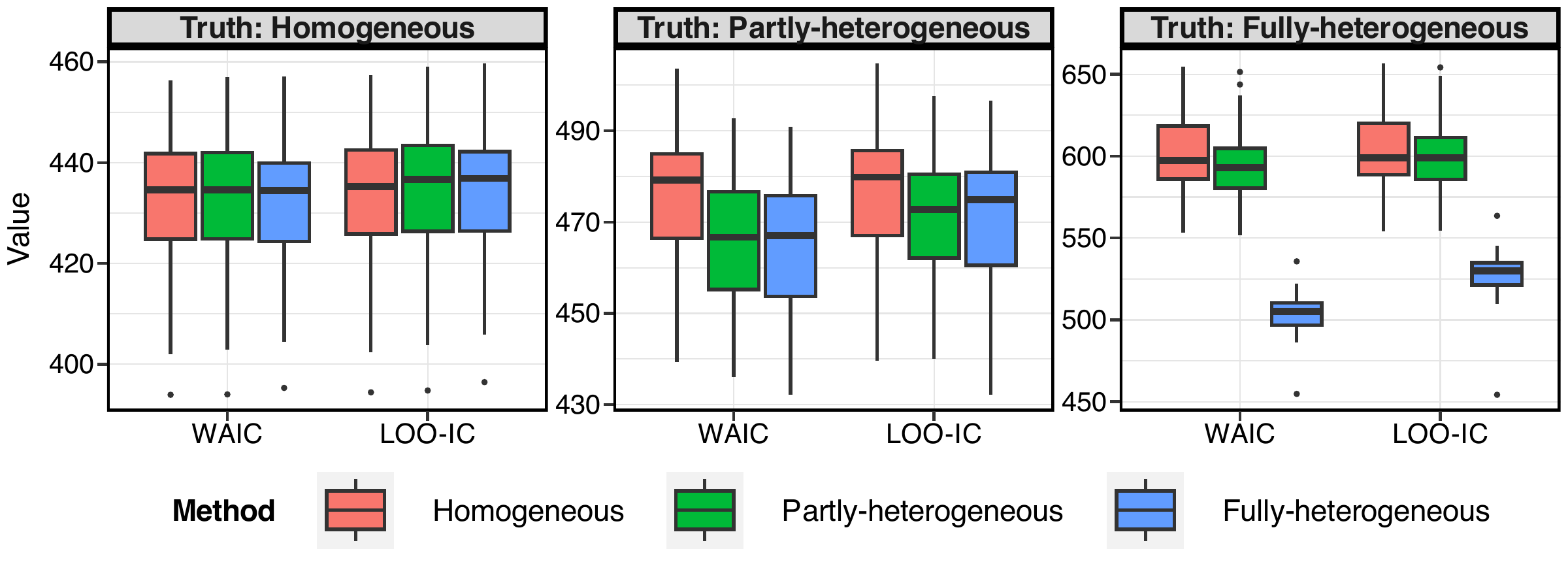}
   \caption{The box-plots of widely applicable information criterion (WAIC) and Leave-one-out cross-validation information criterion (LOO-IC) from 50 replications. Panels indicate true simulation scenarios. Lower values are better.}\label{fig: simulation ic}
\end{figure}
Figure~\ref{fig: simulation ic} presents the model comparison metrics in different true scenarios using the \textit{widely applicable information criterion (WAIC)} and \textit{Leave-one-out cross-validation information criterion (LOO-IC)}. For all methods, Figure~S5 in the supplement compares \textit{in} and \textit{out-of-sample} performances averaged over MITS-VA cause pairs for all countries. Summarizing all figures we find that all methods perform similarly when the misclassification matrix is truly homogeneous. However, where sensitivities exhibit heterogeneity, the heterogeneous methods perform similarly to each other, and they both surpass the performance of the homogeneous model. Finally, in cases where the misclassification matrix is fully-heterogeneous, the fully-heterogeneous method significantly outperforms both the homogeneous and partly-heterogeneous methods. This highlights the fully-heterogeneous method's flexibility in adapting to different situations and demonstrates its efficacy in capturing heterogeneity. %This highlights the efficacy of the fully-heterogeneous method in flexibly adapting to different situations and demonstrating its superior performance in capturing and quantifying heterogeneity.

% \begin{figure}
%      \centering
%      \begin{subfigure}[b]{\textwidth}
%          \centering
%          \includegraphics[width=\textwidth]{hetsim-avgmse.pdf}
%          \caption{}
%          \label{subfig: simulation avgmse}
%      \end{subfigure}
%      \vfill
%      \begin{subfigure}[b]{\textwidth}
%          \centering
%          \includegraphics[width=\textwidth]{hetsim-avgintscore-insample.pdf}
%          \caption{}
%          \label{subfig: simulation avgintscore insample}
%      \end{subfigure}
%      \vfill
%      \begin{subfigure}[b]{\textwidth}
%          \centering
%          \includegraphics[width=\textwidth]{hetsim-avgintscore-outofsample.pdf}
%          \caption{}
%          \label{subfig: simulation avgintscore outofsample}
%      \end{subfigure}
%         \caption{{\bf Performance in simulated data.} Columns represent different true simulation scenarios. For 3 methods (coded in different colors), Figures (\ref{subfig: simulation avgmse})--(\ref{subfig: simulation avgintscore insample}) compare their \textit{in-sample} performance whereas (\ref{subfig: simulation avgintscore outofsample}) compare their \textit{out-of-sample} performance. For each method in each true scenario, it presents boxplots of \textit{widely applicable information criterion (WAIC)} and \textit{Leave-one-out cross-validation information criterion (LOO-IC)} over 50 replications.}
%         \label{fig: simulation performance}
% \end{figure}

\section{Country-resolved Misclassification Rates of InSilicoVA}\label{sec: Misclassification Analysis for InsilicoVA in CHAMPS}

Here we revisit the CHAMPS data introduced in Section~\ref{sec: Motivating Dataset From Child Health and Mortality Prevention Surveillance (CHAMPS) Network} and implement the proposed country-specific model to estimate InSilicoVA's misclassification rates for neonatal deaths. 

\begin{figure}[!t] %  figure placement: here, top, bottom, or page
   \centering
   \includegraphics[width=\textwidth]{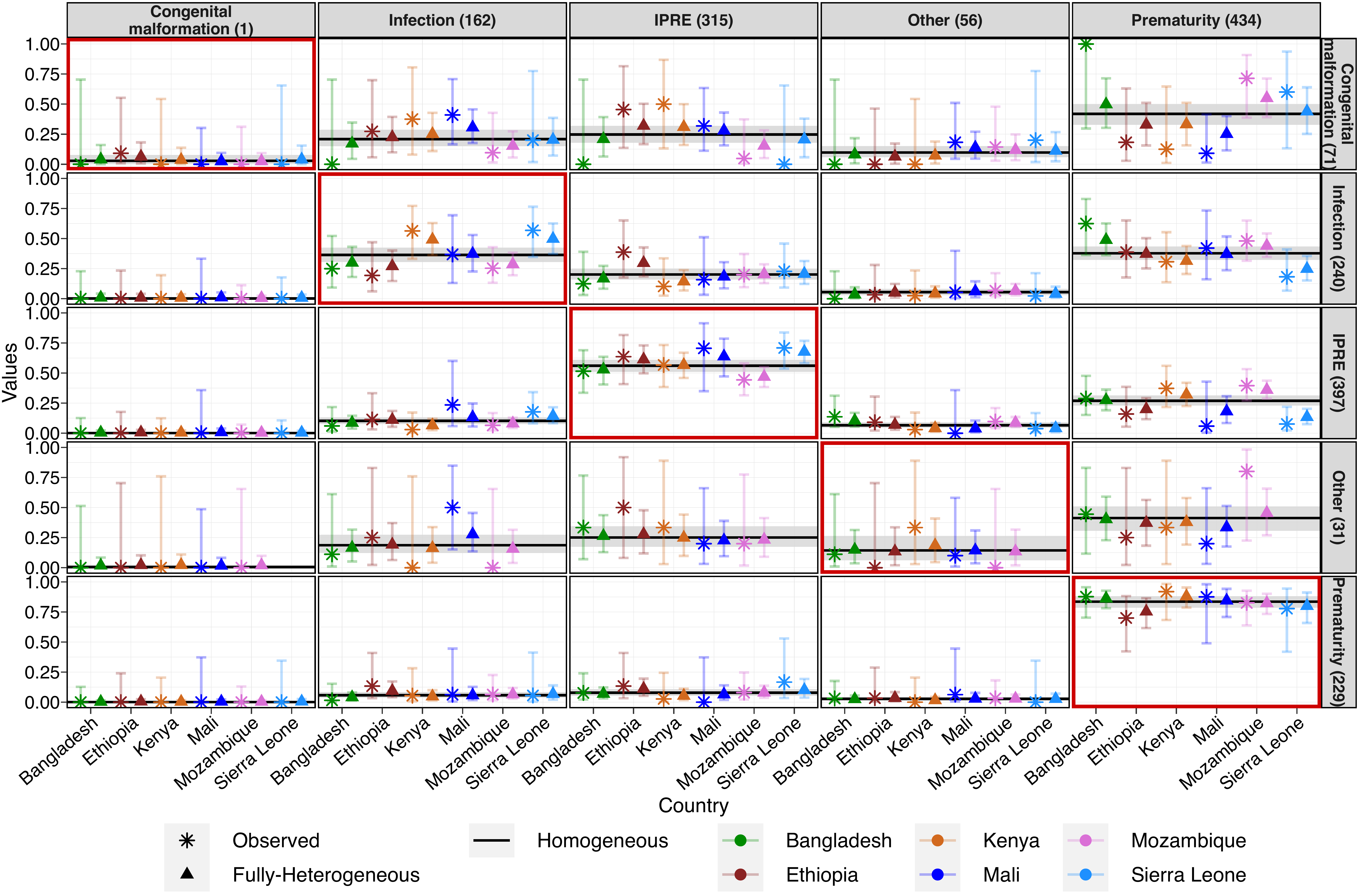}
   \caption{Comparison of observed misclassification rates with in-sample point (posterior mean) and uncertainty (95\% credible interval) estimates from homogeneous and fully-heterogeneous models. Rows and columns refer to MITS and VA-predicted causes.}
   \label{fig: insample classification estimates}
\end{figure}
\subsection{Misclassification Rate Estimates}\label{sec:insample}

%Following the model selection performance, 
We implemented the three methods as outlined in Section~\ref{sec: Simulation Study} but focus our comparison on homogeneous and fully-heterogeneous methods for brevity, unless stated otherwise. 
In Figure~\ref{fig: insample classification estimates} we present the point (posterior mean) and uncertainty estimates (95\% credible interval) of country-specific misclassification rates from the two methods and compare them with the observed data. When strong evidence of heterogeneity is absent in observed rates, the estimates from fully-heterogeneous method closely resemble the homogeneous estimates. Examples include sensitivity and false positives for MITS cause prematurity. On the other hand, if there is a substantial heterogeneity in observed rates, the fully-heterogeneous estimates better match the observed rates as compared to homogeneous estimates. This is seen for sensitivities of IPRE and false positive rates of IPRE-prematurity in Mozambique and Sierra Leone. Collectively, the fully-heterogeneous model yields more accurate in-sample estimates in the presence of heterogeneity, while demonstrating comparable performance to the homogeneous model in situations where heterogeneity is relatively insignificant. %Taken together, the results indicate that the fully-heterogeneous model provides more accurate in-sample estimates in the presence of heterogeneity while delivering similar performance to the homogeneous model when heterogeneity is not a significant factor.

\begin{figure}[!t]
     \centering
         \includegraphics[width=\textwidth]{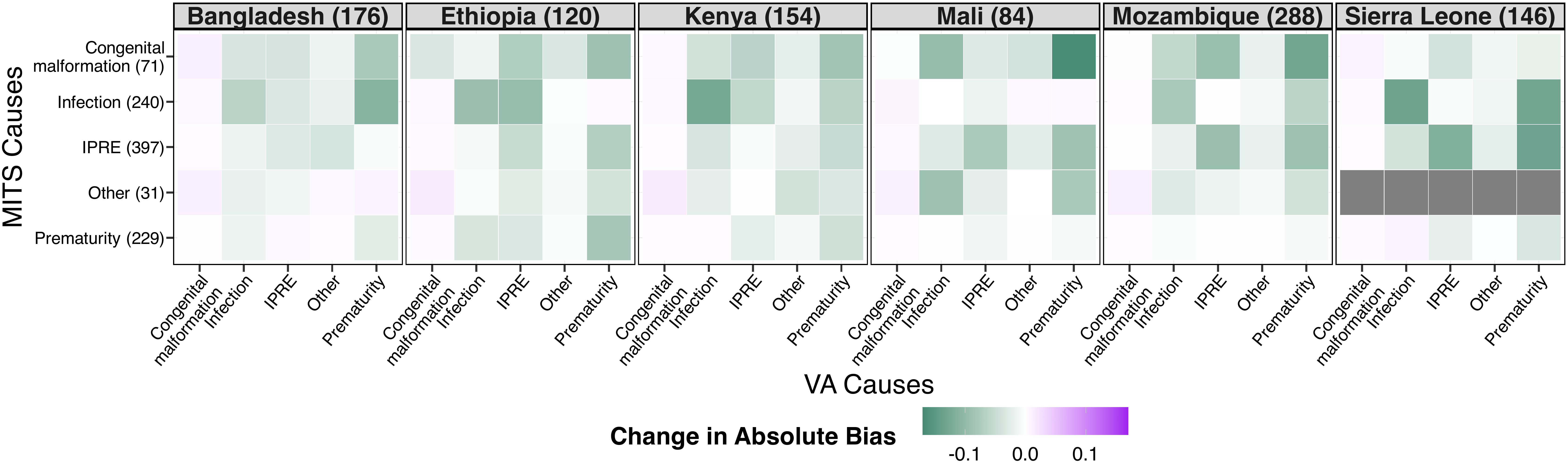}
         \includegraphics[width=\textwidth]{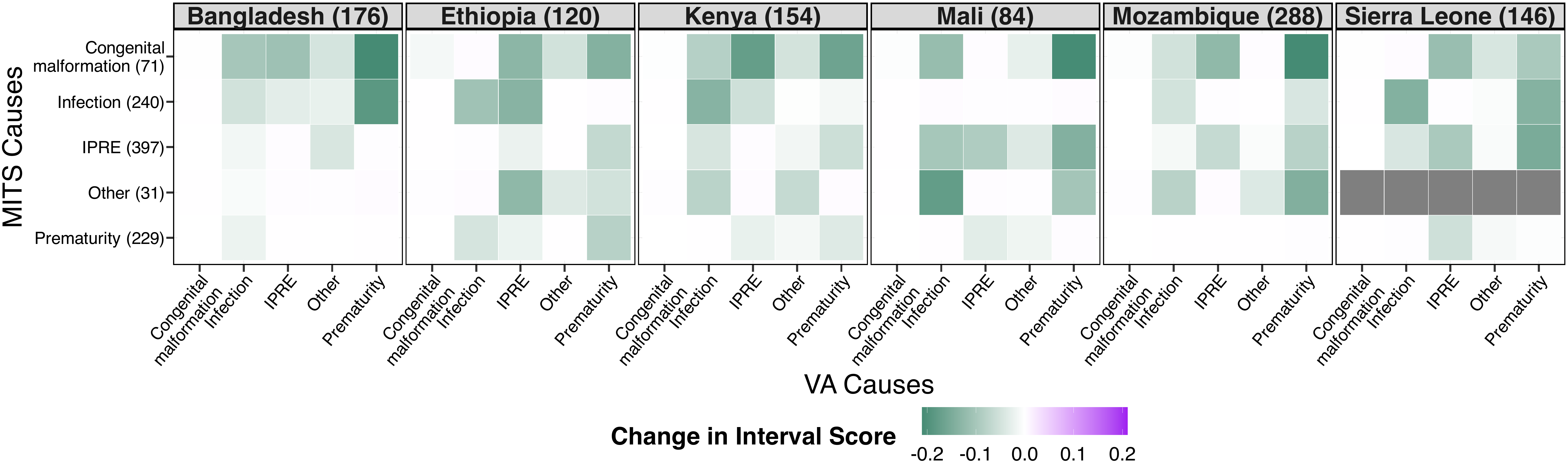}
        \caption{The figure compares the change in absolute bias in posterior means (\textit{top panel}) and interval scores of 95\% credible intervals (\textit{bottom panel}) in the fully-heterogeneous method compared to the homogeneous method. Rows and columns are MITS and VA causes. Negative values are better (green color)  indicating reduction in bias / interval score.}
        \label{fig: insample bias, interval score reduction}
\end{figure}
\subsection{Model Comparison}\label{sec:metrics}
%We applied all three models outlined in Section~\ref{sec: Simulation Study} to the dataset and assessed their performance in terms of model selection, in-sample performance on the training data, and leave-one-country out-of-sample performance. The results are presented below.xx
%\noindent {\bf \textit{Model comparison.}} 
Figure~\ref{fig: insample bias, interval score reduction} presents the model comparison metrics between the fully-heterogeneous model and the homogeneous model for both point and interval estimation. Point estimates are compared using absolute bias in posterior means and 95\% credible intervals are evaluated using interval scores \citep{Gneiting2007}. %of the  between the fully-heterogeneous method and the homogeneous method. 
The results overwhelmingly favor the heterogeneous model. Combined across countries, it reduced bias for $73\%$ cause pairs. For half of the cause pairs the bias decreased by at least 44\%, with reductions as high as 98\% (MITS cause congenital malformation, VA cause prematurity in Mali). Only for the first columns of the misclassification matrices (corresponding to some MITS cause, and VA cause congenital malformation), the homogeneous model has slightly less bias. This is because InSilicoVA does not predict congenital malformation, so the observed misclassification rates in this column are exactly 0. The homogeneous model on account of pooling data is less reliant on the prior and estimates are also nearly zero. For the heterogenous model, the estimates are also very close to zero but there is still bias on account of being more influenced by the prior which gives some mass away from zero. %\pink{AD: I added quite a bit here. Please check if this is accurate.} 

The heterogeneous model also reduced interval scores for $57\%$ cause pairs across countries. There was at least $87\%$ reduction in scores for $50\%$ cause pairs, with reductions as high as $98\%$. 
Additionally, Figure~S7 in the supplement displays the LOO-IC and WAIC for all three methods. Compared to the homogeneous method, the partly-heterogeneous method enhances the information criteria by 6--7\%. However, the fully-heterogeneous method outperforms all, improving the criteria by 5--8\% over partly-heterogeneous and an impressive 11--15\% over the homogeneous method. In summary, the fully-heterogeneous model produces more accurate in-sample estimates than the homogeneous model. %These findings collectively suggest that the fully-heterogeneous model estimates more accurate in-sample estimates compared to the homogeneous model. 

%\medskip
%\noindent {\bf \textit{In-sample performance.}}
%\paragraph{In-sample performance.} 

\begin{figure}[!t] %  figure placement: here, top, bottom, or page
   \centering
   \includegraphics[width=\textwidth]{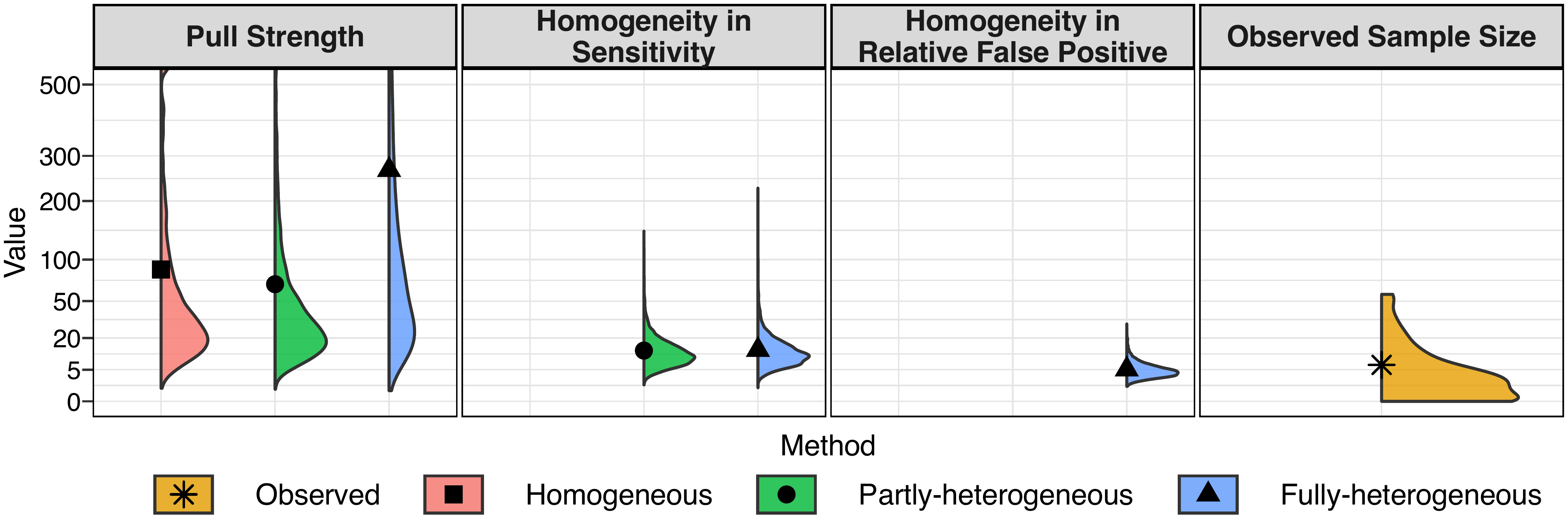}
   \caption{Three panels on the left show posteriors of effect sizes from three models. Larger values indicate a stronger effect. The last panel displays the frequency distribution of sample sizes observed across cause pairs and countries for the comparative analysis of effect size magnitude. Black points denote means of respective distributions. \textit{For clear representation, the y-axis is plotted on the square root scale and is restricted to the range of means.}}
   \label{fig: effect size}
\end{figure}
\subsection{Estimates of Effect Size, Intrinsic Accuracy, and Pull}

{\bf \em Estimates of Effect Sizes $\omega_P$, $\omega_S$, $\omega_R$.} We discuss estimates of key modeling parameters that define the proposed framework. Figure~\ref{fig: effect size} compares posterior densities of effect size estimates from three different models. With effect sizes carrying the interpretation of prior sample sizes per cause or category, the estimates must be compared in conjunction with sample sizes observed across cause pairs and countries (Figure~\ref{fig: effect size}, last panel). 

First, we note that the effect size estimates for parameters common to the models are generally consistent with each other. For example, the posterior density pull strength $\omega_P$, which dictates the shrinkage towards the base model in all 3 models, have roughly similar shapes and modes. This is also seen for $\omega_S$ controlling homogeneity in the sensitivity. 
This underscores the validity of the adaptive nested structure of the framework. Second, the estimates of $\omega_P$ are significantly higher (posterior means range 75--400) than the maximum observed sample size (57). This indicates very strong evidence towards the base model structure indicating systematic preference underlying InSilicoVA's misclassification mechanism. Third, the estimate of $\omega_S$ is significantly lower (posterior means $\approx 12$) compared to that of pull strength and the maximum sample size. This indicates a strong presence of heterogeneity in sensitivities. Finally, the estimates for the degree of homogeneity in relative FP ($\omega_R$) is even lower (posterior mean $\approx 5$) than $\omega_S$. This implies significant heterogeneity in InSilicoVA's relative FP across countries, with a more pronounced degree compared to that in sensitivity. Together, this suggests a strong systematic preference and a pronounced presence of heterogeneity across different countries in InSilicoVA's misclassification rates.

\begin{figure}[!h]
     \centering
          \begin{subfigure}[b]{0.495\textwidth}
         \centering
         \includegraphics[width=\textwidth]{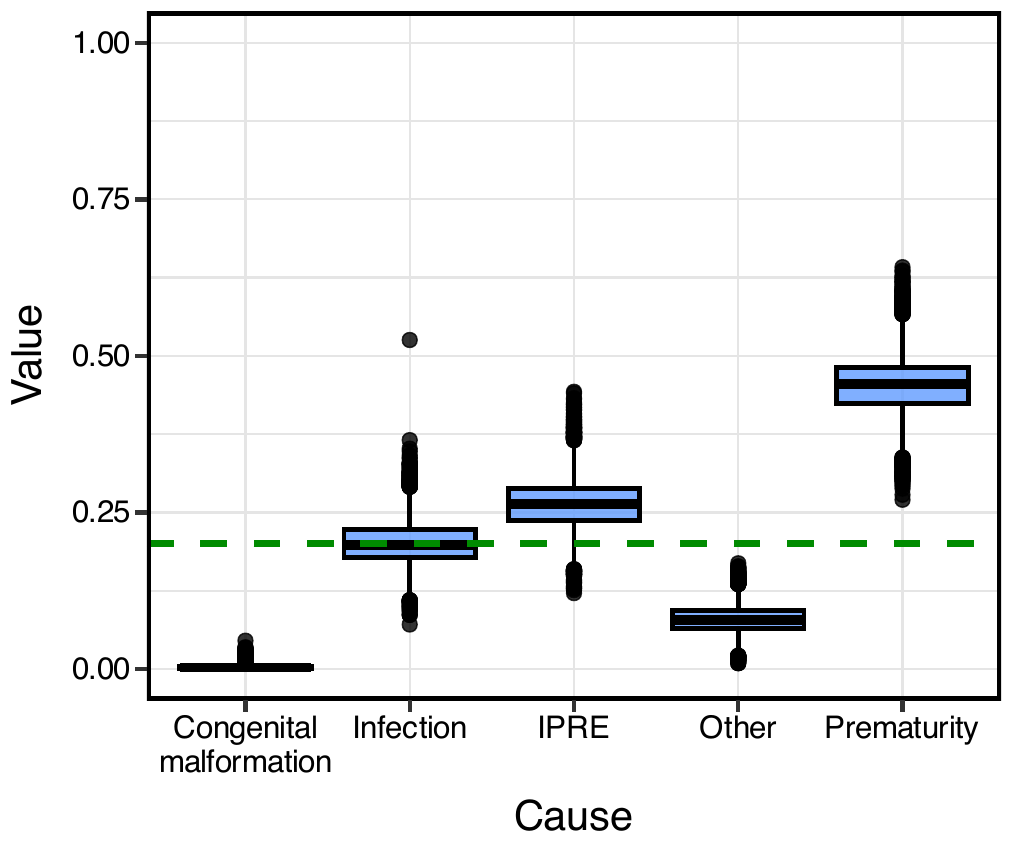}
         \caption{Pull}
         \label{subfig: single-neonate-insilicova-pull}
     \end{subfigure}
     \hfill
     \begin{subfigure}[b]{0.495\textwidth}
         \centering
         \includegraphics[width=\textwidth]{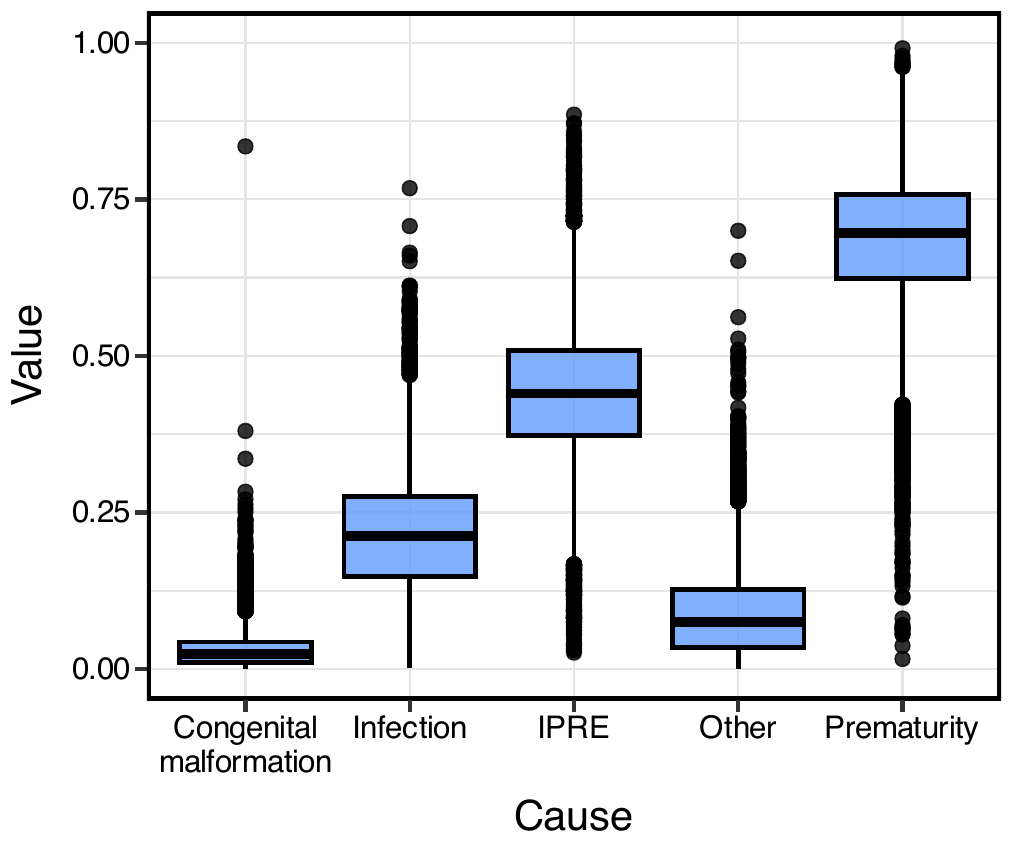}
         \caption{Intrinsic accuracy}
         \label{subfig: single-neonate-insilicova-intrinsic_accuracy}
     \end{subfigure}
        \caption{The posterior of intrinsic accuracies and pull for InSilicoVA among neonatal deaths in CHAMPS. The horizontal dashed line in (b) corresponds to no preference with a uniform pull of 1/5 for each cause.}
        \label{fig: single-neonate-insilicova intrinsic_accuracy & pull}
\end{figure}
\noindent {\bf \textit{Estimates of Intrinsic Accuracy $\boldsymbol{(a)}$ and Pull $\boldsymbol{(\alpha)}$.}}
%\paragraph{Estimates of intrinsic accuracy and pull.} 
Figure~\ref{fig: single-neonate-insilicova intrinsic_accuracy & pull} showcases the posteriors of intrinsic accuracies $a_i$'s and pulls $\alpha_j$'s from the fully-heterogeneous method, and they offer two key findings. Note that, in the absence of systematic preference, an algorithm would uniformly misclassify all causes, implying an {\em average pull} of 1/5 (green dashed line) for each cause. %Contrary to this, Figure~\ref{subfig: single-neonate-insilicova-pull} shows that pull estimates for congenital malformation, prematurity, and other causes significantly differ from it, clearly demonstrating a presence of systematic preference. It overstates Prematurity and suppresses Congenital malformation and Other. 
Contrary to this, the pull estimates in Figure~\ref{subfig: single-neonate-insilicova-pull} suggest that the algorithm overstates prematurity and suppresses congenital malformation and other, clearly demonstrating a presence of systematic preference. We display intrinsic accuracy estimates for all causes in Figure~\ref{subfig: single-neonate-insilicova-intrinsic_accuracy}. While observed sensitivities are induced by both intrinsic accuracy and pull, the figure shows that intrinsic accuracy dominates the influence for most causes, with prematurity having the highest intrinsic accuracy, followed by IPRE. 
%While observed sensitivities are a combination of intrinsic accuracy and pull, the estimates in Figure~\ref{subfig: single-neonate-insilicova-intrinsic_accuracy} indicate that, for most causes, intrinsic accuracy is the primary factor influencing sensitivity predictions, with prematurity having the highest intrinsic accuracy, followed by IPRE. 
%While the observed sensitivities result from a combination of intrinsic accuracy and pull, the estimates in Figure~\ref{subfig: single-neonate-insilicova-intrinsic_accuracy} indicate that, for most causes, the main factor contributing to the prediction of sensitivities is their intrinsic accuracy. The intrinsic accuracy for prematurity is the highest, with IPRE coming second. %Also, there is a strong correlation between intrinsic accuracy and pull, which is on par with what happens when designing algorithms by balancing accuracy and error. This is also observed for Type I error and power in designing statistical hypothesis tests.

%\medskip
\subsection{Extrapolation Performance}\label{sec:extra}
%\paragraph{Out-of-sample performance.}
For out-of-sample comparison, we train methods by omitting one country at a time and assess predictive accuracy with respect to observed rates in the excluded country. Figure~S6 in the supplement compares out-of-sample point (predictive mean) and uncertainty predictions (95\% prediction interval) from homogeneous and fully-heterogeneous methods. %, where the respective country was not used in training.  %We carry this out by iteratively excluding each country one by one. 
The point predictions from the two methods are similar and it conforms to the discussion in Section~\ref{sec:pred}. %This is because the prediction for a new country is centered around its estimate from the homogeneous model while country-specific random effects account for heterogeneity through predictive uncertainties (see Section \ref{sec:pred}). % by accounting for uncertainty, we assess interval scores of 95\% prediction intervals for performance evaluation.
%In Figure~\ref{sfig: outofsample classification estimates}, we compare out-of-sample point (predictive mean) and uncertainty predictions (95\% prediction interval) from homogeneous and fully-heterogeneous methods, where the respective country was not used in training. 
The homogeneous model provides narrower intervals compared to the fully-heterogeneous model as it ignores heterogeneity. The intervals from the fully-heterogeneous method account for heterogeneity and are more accurate. This performance difference is evident in the sensitivity of infection and false positive rates of IPRE-prematurity.

\begin{figure}[!h] %  figure placement: here, top, bottom, or page
   \centering
   \includegraphics[width=\textwidth]{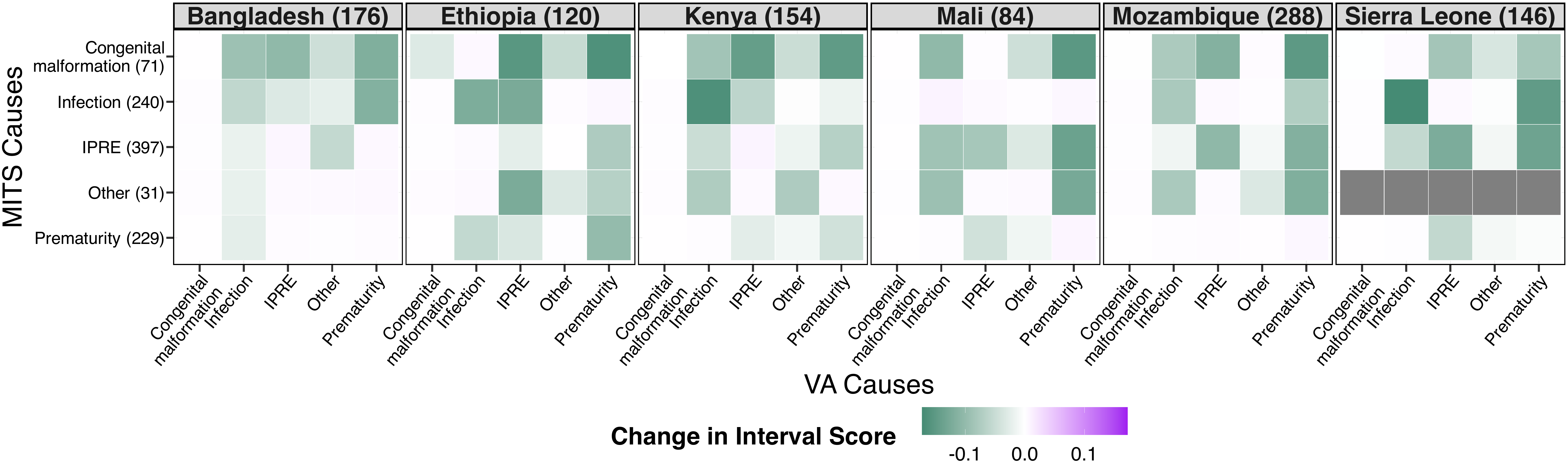}
   \caption{The figure compares differences in interval scores of 95\% prediction intervals in the fully-heterogeneous model compared to the homogeneous. Rows and columns are MITS and VA causes. Smaller values are better (green color), indicating more reduction in scores.}
   \label{fig: outofsample interval score reduction}
\end{figure}
% \begin{figure}[!t] %  figure placement: here, top, bottom, or page
%    \centering
%    \includegraphics[width=\textwidth]{figures/single-neonate-insilicova-outofsample-mu.pdf}
%    \caption{Comparison of observed misclassification rates with out-of-sample point (predictive mean) and uncertainty (95\% prediction interval) predictions from homogeneous and fully-heterogeneous methods. Rows and columns refer to MITS and VA-predicted causes. \pink{AD: May need to move this to Supplement as we are half page over currently.}}
%    \label{sfig: outofsample classification estimates}
% \end{figure}

Figure~\ref{fig: outofsample interval score reduction} formally compares the uncertainty quantification. Compared to the homogeneous method, the fully-heterogeneous model reduces the interval scores for $55\%$ cause pairs across countries. There is at least $77\%$ reduction for half of the cause pairs, with reductions as high as $94\%$. This collectively demonstrates the advantages of the fully-heterogeneous method in offering more accurate out-of-sample predictions. %This collectively demonstrates the benefits of the fully-heterogeneous method in providing more precise out-of-sample predictions by effectively accounting for heterogeneity.

\begin{figure}[!b] %  figure placement: here, top, bottom, or page
   \centering
   \includegraphics[width=\textwidth]{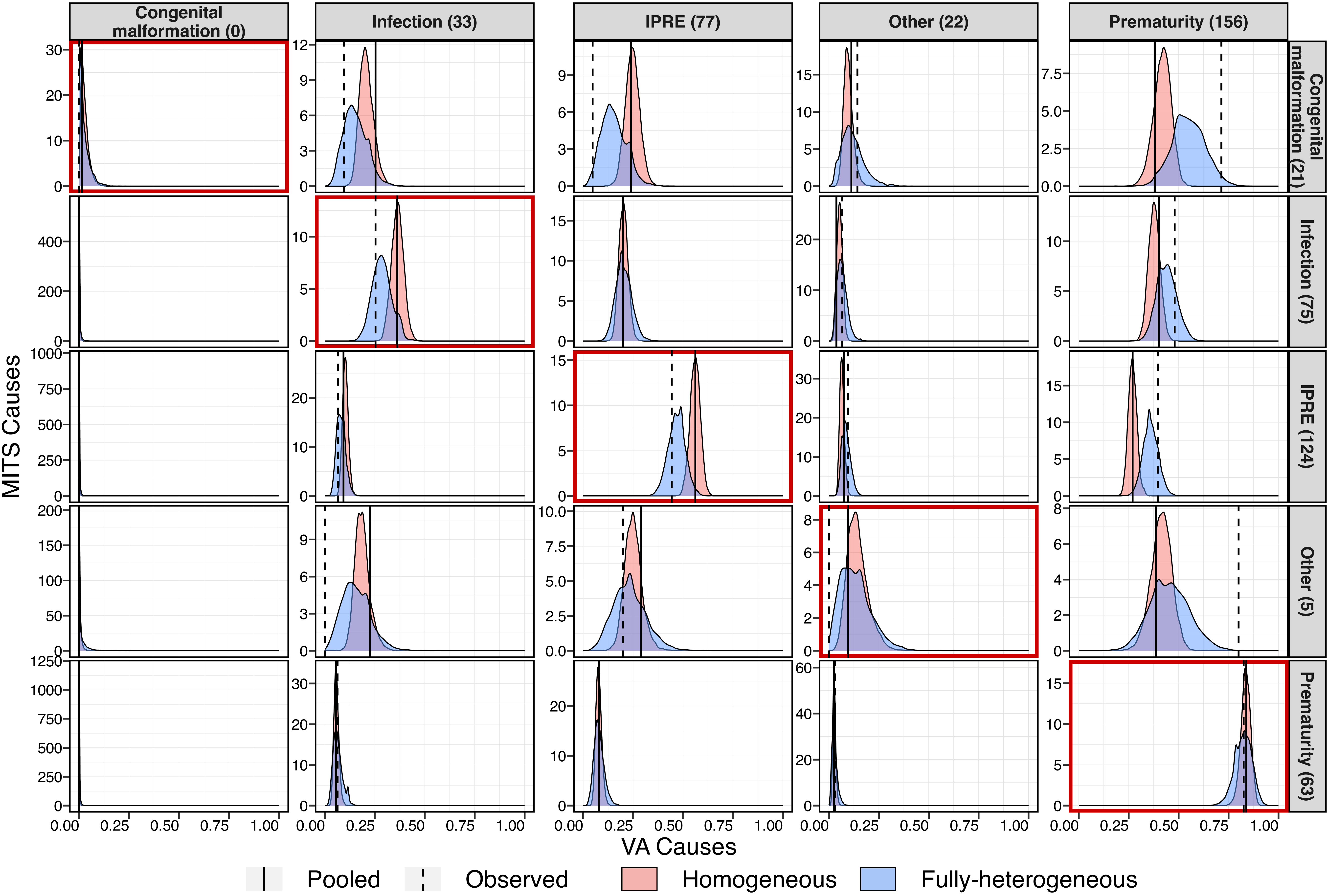}
   \caption{Comparison of misclassification rate estimates in Mozambique from homogeneous and fully-heterogeneous models. The solid vertical lines denote the proportions pooled across countries while the dashed lines denote that observed in Mozambique. The numbers enclosed in panel headings are sample sizes in Mozambique.}
   \label{fig: mozambique}
\end{figure}
\subsection{Misclassification Matrix Estimate in Mozambique}

%\paragraph{Misclassification matrix estimate in Mozambique.} 
%A future application of the research presented here would be to leverage the  misclassification matrix estimate for Mozambique to perform calibration of the COMSA-MZ data and produce calibrated CSMF estimates with improved validity. This will update the current calibrated estimates which used a homogenous misclassification matrix pooling data across all CHAMPS countries \citep{Fiksel2023,Gilbert2023}. 
% \cite{Fiksel2023} and \cite{Gilbert2023} calibrated VA data in Mozambique using the misclassification matrix estimated from CHAMPS data pooled across all countries under a homogeneity assumption.

Assuming homogeneous misclassification rates, \cite{Fiksel2023} and \cite{Gilbert2023} estimated Mozambique's misclassification matrix by pooling CHAMPS data from all countries.
Figure~\ref{fig: mozambique} contrasts the uncertainty quantified posterior estimates of misclassification rates in Mozambique derived from our homogeneous (red) and fully-heterogeneous (blue) models.
%In Figure~\ref{fig: mozambique} we compare the uncertainty quantified posterior estimates of misclassification rates in Mozambique obtained from our homogeneous (red) and fully-heterogeneous (blue) models. 
When for an MITS-VA cause pair the empirical misclassification rate in Mozambique (dashed line) significantly differs from the pooled empirical rate (solid line), and an adequate number of samples is available, the fully-heterogeneous method provides a more precise estimate.
%When the empirical misclassification for Mozambique (dashed line) for an MITS-VA cause pair significantly deviates from the pooled empirical rate (solid line), and there are enough samples available, the posterior from the fully-heterogeneous method offers a more precise estimate. 
This is evident in sensitivities for infection and IPRE, and false positives for congenital malformation-infection and IPRE-prematurity. %In other scenarios, both posteriors are closer to the pooled proportion. 
When sample sizes are insufficient, this approach enhances the stability of country-specific estimates by shrinking them towards the pooled estimate. %while providing equally accurate estimates otherwise. 
This underscores the adaptive nature of the proposed framework in providing more representative country-specific estimates in the presence of enough evidence while shrinking towards the pooled estimate otherwise to improve precision. % in absen improves the validity of country-specific estimates by effectively leveraging local-global  misclassification patterns.

\section{Discussion}\label{sec: Discussion}

Accurate misclassification rate estimation for VA algorithms is pivotal in obtaining valid CSMF estimates based on VA-calibration. The labeled MITS-VA COD data collected in CHAMPS is a vital resource in this context. Although prior research assumes the same misclassification rates for all countries, our findings expose substantial local heterogeneity and emphasize the need for addressing it. With the number of parameters growing as a quadratic in the number of causes and linearly in the number of countries, the limited availability of MITS cases poses a challenge in obtaining meaningful country-specific estimates.

We introduced a novel framework for heterogeneous modeling of misclassification matrices with three key components: $(i)$ the base model to account for the systematic preference of VA algorithms, $(ii)$ the exact error matrix decomposition into sensitivities and relative FP to generalize the base model, and $(iii)$ the hierarchically-defined nested Bayesian framework to incorporate heterogeneity. The base model introduces the concept of pull to quantify systematic preferences for certain causes. Theorem \ref{thm: logodds characterization} characterizes its manifestation in terms of constant misclassification odds. The nested structure enables the framework to systematically expand from a simple homogeneous model to a complex country-specific model in a data-driven manner. In the absence of evidence, the shrinkage prior on effect sizes encourages lower complexity effectively compensating for limited MITS cases.

Findings from this research have immediate and future implications for global health. %that point toward future research directions. 
%In the short-term, these models are being used to generate and publish calibrated CSMF estimates for Mozambique by using the Mozambique-specific misclassification matrix estimate estimated in this study with the VA-only data collected in COMSA-Mz. 
In the short term, the Mozambique-specific misclassification matrix estimated here will be combined with VA-only data collected in COMSA-Mz to produce calibrated CSMF estimates. For different VA algorithms and age groups, we will perform comparative analyses of misclassification rates and resulting CSMF estimates. In the longer term, these estimates will be used to calibrate VA data from 100 studies for the next set of regional and national under-5 CSMF estimates for the Sustainable Development Goals \citep{perin2022global}.

In numerous scientific challenges, the mechanisms behind data generation are either unknown or cannot be guaranteed. The notion of misclassification thus pervades various domains in Statistics, extending much beyond cause-of-death analyses. Statistical inference for association studies using \textit{electronic health records (EHR)} is an example in health-related research where outcomes are often misclassified \citep{Beesley2022,sinnott2014}. More recently, we are seeing rapid advancements in \textit{large language models} and other computer-coded algorithms across \textit{machine learning} and \textit{artificial intelligence} that do not perfectly capture truth. In this regard, \textit{transfer learning} has become increasingly relevant which aims at improving a \textit{target learner}'s performance on \textit{target domains} by transferring the knowledge contained in different but related \textit{source domains} \citep{Wang2018,Zhuang2021}. Given the inherent nature of misclassification, it timely presents opportunities for us to extend the fundamental concepts presented in this research to diverse scientific problems. Instances such as these, among many others, underscore that the crux of the addressed issue resides at the intersection of multiple active and promising scientific problems. Exploring and expanding these concepts, as well as delving into other facets to address them, presents an intriguing avenue for future research.

\section*{Supplementary Materials} 
Proofs of theoretical results, simulation details, and additional figures that generate country-specific misclassification matrix estimates are provided in the Supplementary Materials. Codes will be added to the existing \textit{VA-calibration} \texttt{R} package on CRAN.
% Proofs of theoretical results, details of simulation studies, and additional figures pertaining to the analysis that generates estimates of country-specific misclassification matrices are provided in the Supplementary Materials file.

\section*{Acknowledgement}
S.P., S.Z., A.D. were supported by the Bill and Melinda Gates Foundation Grant INV-034842. D.B. received support from the Bill and Melinda Gates Foundation Grant OPP1126780. %\pink{AD: Dianna please add any funding support that needs to be acknowledged.} 
The authors are grateful to  Emily Wilson and Dr. Henry Kalter for processing the CHAMPS data. The findings and conclusions in this report are those of the authors and do not necessarily represent the views of the US Centers for Disease Control and Prevention.

%From a methodological standpoint, some VA algorithms are \textit{probabilistic} in nature and generate \textit{multi-cause} COD. We will extend the framework to accommodate this. Furthermore, VA algorithms offer COD predictions at a more granular level of causes. With the limited availability of MITS cases, obtaining stable country-specific estimates for this finer set of causes is presently unfeasible. We will investigate to expand the framework to accommodate causes at a higher resolution.

\bibliographystyle{apalike}
\bibliography{Bibliography-MM-MC.bib}

\end{document}

\maketitle

\begin{abstract}
In this supplementary file, we provide some technical details and additional materials for the main article.
\end{abstract}

\section{Proof of Theorem~3.1 in The Main Article}

\paragraph{Proof of \textit{`if'} part.} Let us fix an MITS cause $i$ arbitrarily. The constant odds condition from Theorem~3.1 in the main article implies that, for some positive $\eta_{jk}$ for $j \neq k$,
\begin{equation*}
    \frac{\phi_{ij}}{\phi_{ik}} = \eta_{jk}, \quad \forall i \text{ and } j \neq k \neq i.
\end{equation*}
Being odds, $\eta_{jk}$'s also satisfy
\begin{equation*}
    \eta_{jk} \eta_{kl} = \eta_{jl}, \quad \forall j\neq k \neq l.
\end{equation*}
Setting $l=C$ without loss of generality we get
\begin{equation*}
    \eta_{jk} = \frac{\eta_{jC}}{\eta_{kC}} = \frac{\theta_j}{\theta_k}, \quad \forall j\neq k,
\end{equation*}
where $\theta_j$'s are positive. Combining this with the decomposition (4) in the main article we get
\begin{equation*}
    \frac{\phi_{ij}}{\phi_{ik}} = \frac{q_{ij}}{q_{ik}} = \frac{\theta_j}{\theta_k} \quad \Rightarrow \quad q_{ij} = \frac{\theta_j}{\theta_k} q_{ik}, \quad \forall i \text{ and } j \neq k \neq i.
\end{equation*}
Using $\sum_{j \neq i} q_{ij} = 1$, we obtain
\begin{equation*}
    q_{ij} = \frac{\theta_j}{\sum_{l \neq i} \theta_l}, \quad \forall i \text{ and } j \neq i.
\end{equation*}
Defining the pull as $\alpha_j = \theta_j / \sum_l \theta_l$ completes the proof.

\paragraph{Proof of \textit{`only if'} part.} Recall from (5) in the main article that, the relative FP $q_{ij}$ in the base model equals to $\alpha_j / (1-\alpha_i)$. Together with the decomposition (4), this yields
\begin{equation*}
    \frac{\phi_{ij}}{\phi_{ik}} = \frac{\alpha_j}{\alpha_k}, \quad \forall j,k \neq i.
\end{equation*}
Defining $\alpha_j = \theta_j$ completes the proof.

\section{Proof of Proposition~3.1 in The Main Article}

According to (1) in the main article, the classification probabilities in the base model are
\begin{align*}
    \phi_{ii} &= a_i + (1-a_i)\alpha_i \quad  \forall i,\\
    \phi_{ij} &= (1-a_i)\alpha_j \quad \forall j \neq i,
\end{align*}
where $\bs{a} = (a_1,\dots,a_C)^\T$ are intrinsic accuracies with $a_i \in (0,1)$, and $\bs{\alpha} = (\alpha_1,\dots,\alpha_C)^\T$ denotes the pull which lies in the $C$--dimensional simplex.

We prove the proposition by contradiction. Suppose, the base model likelihood is not identifiable with respect to $\bs{a}$ and $\bs{\alpha}$. So there exists intrinsic accuracies $\bs{a}_1 \neq \bs{a}_2$, and pulls $\bs{\alpha}_1 \neq \bs{\alpha}_2$ that yields the same likelihood. Mathematically,
\begin{align}
    a_{1i} + (1-a_{1i}) \alpha_{1i} &= a_{2i} + (1-a_{2i}) \alpha_{2i} \quad  \forall i, \nonumber\\
    (1-a_{1i}) \alpha_{1j} &= (1-a_{2i}) \alpha_{2j} \quad \forall j \neq i \label{seq: fp identifiability}.
\end{align}

\noindent Without loss of generality, fix $i=1$. For $j = 2, \dots, C$, (\ref{seq: fp identifiability}) implies
\begin{equation}\label{seq: alpha1 i1}
    \alpha_{1j} = \frac{(1-a_{21})}{(1-a_{11})} \alpha_{2j} = k_1 \alpha_{2j}, \quad \text{where} \quad k_1 = \frac{1-a_{21}}{1-a_{11}}.
\end{equation}
Since $\sum_{j=1}^C \alpha_{1j} = 1$, this implies $\alpha_{11} = 1 - \sum_{j=2}^C \alpha_{1j} = 1 - k_1 (1 - \alpha_{21})$. Fixing $i=2$, we similarly get
\begin{align}
    \alpha_{1j} &= k_2 \alpha_{2j}, \quad \forall j=1,3,\dots,C, \quad \text{with} \quad k_2 = \frac{1-a_{22}}{1-a_{12}}, \label{seq: alpha1j i2}\\
    \alpha_{12} &= 1 - \sum_{j \neq 2} \alpha_{1j} = 1 - k_2 (1 - \alpha_{22}). \label{seq: alpha11 i2}
\end{align}
Equating (\ref{seq: alpha1 i1}) and (\ref{seq: alpha1j i2}) for $j=3$ implies $k_1 = k_2 = k$, say. Finally, equating $\alpha_{12}$ from (\ref{seq: alpha1 i1}) with (\ref{seq: alpha11 i2}) we get
\begin{equation*}
    k \alpha_{22} = 1 - k (1 - \alpha_{22}) \,\, \Rightarrow k=1.
\end{equation*}
This implies $\bs{a}_1 = \bs{a}_2$ and $\bs{\alpha}_1 = \bs{\alpha}_2$.

\begin{figure}[!h] %  figure placement: here, top, bottom, or page
   \centering
   \includegraphics[width=\textwidth]{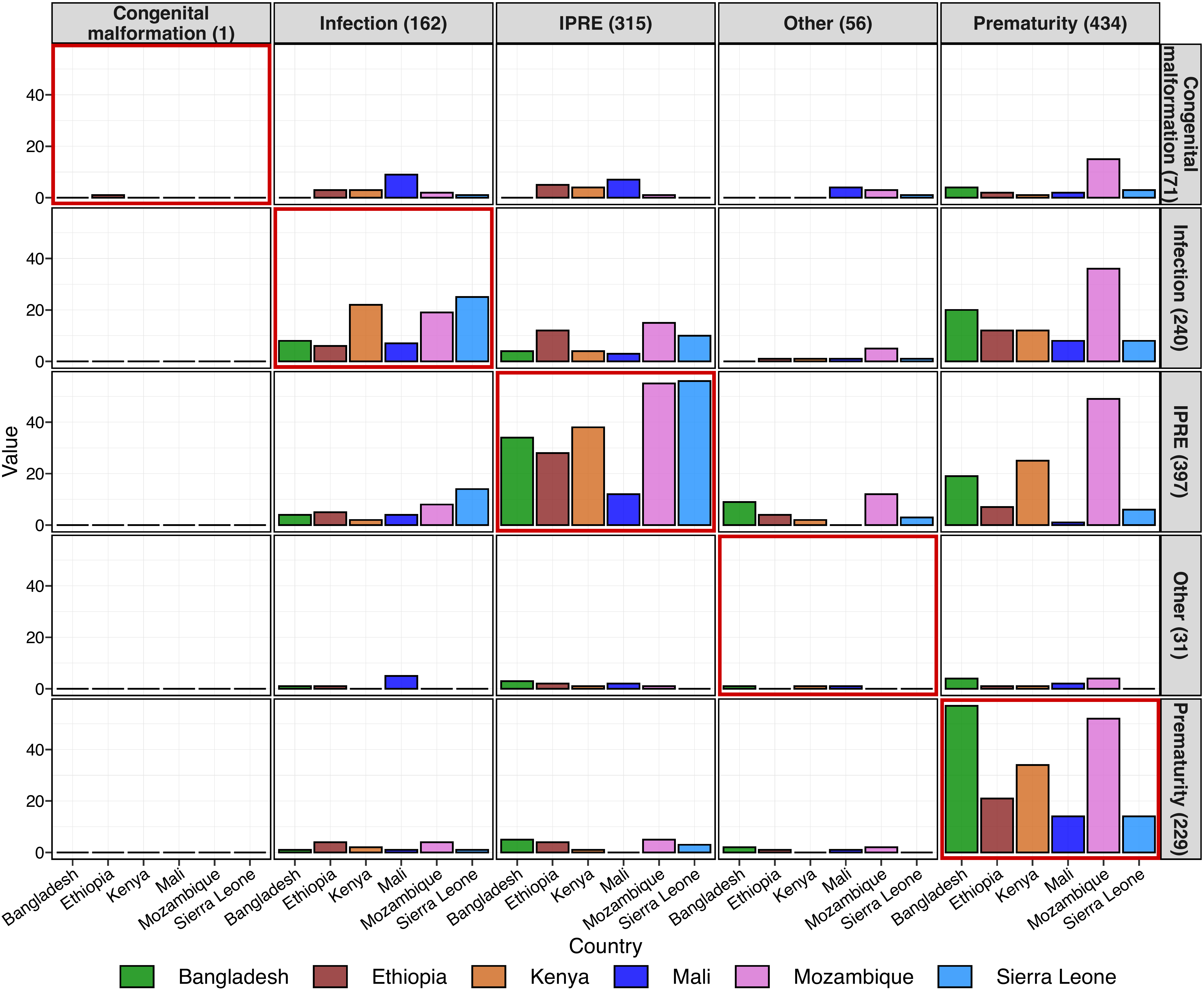}
   \caption{Neonatal deaths observed in the CHAMPS dataset for InSilicoVA across different MITS-VA cause pairs and countries. Rows and columns are MITS and VA-predicted causes. Numbers in parentheses denote the sample size.}
   \label{sfig: observed classification numbers}
\end{figure}

\begin{figure}[!h]
     \centering
     \begin{subfigure}[b]{0.495\textwidth}
         \centering
         \includegraphics[width=\textwidth]{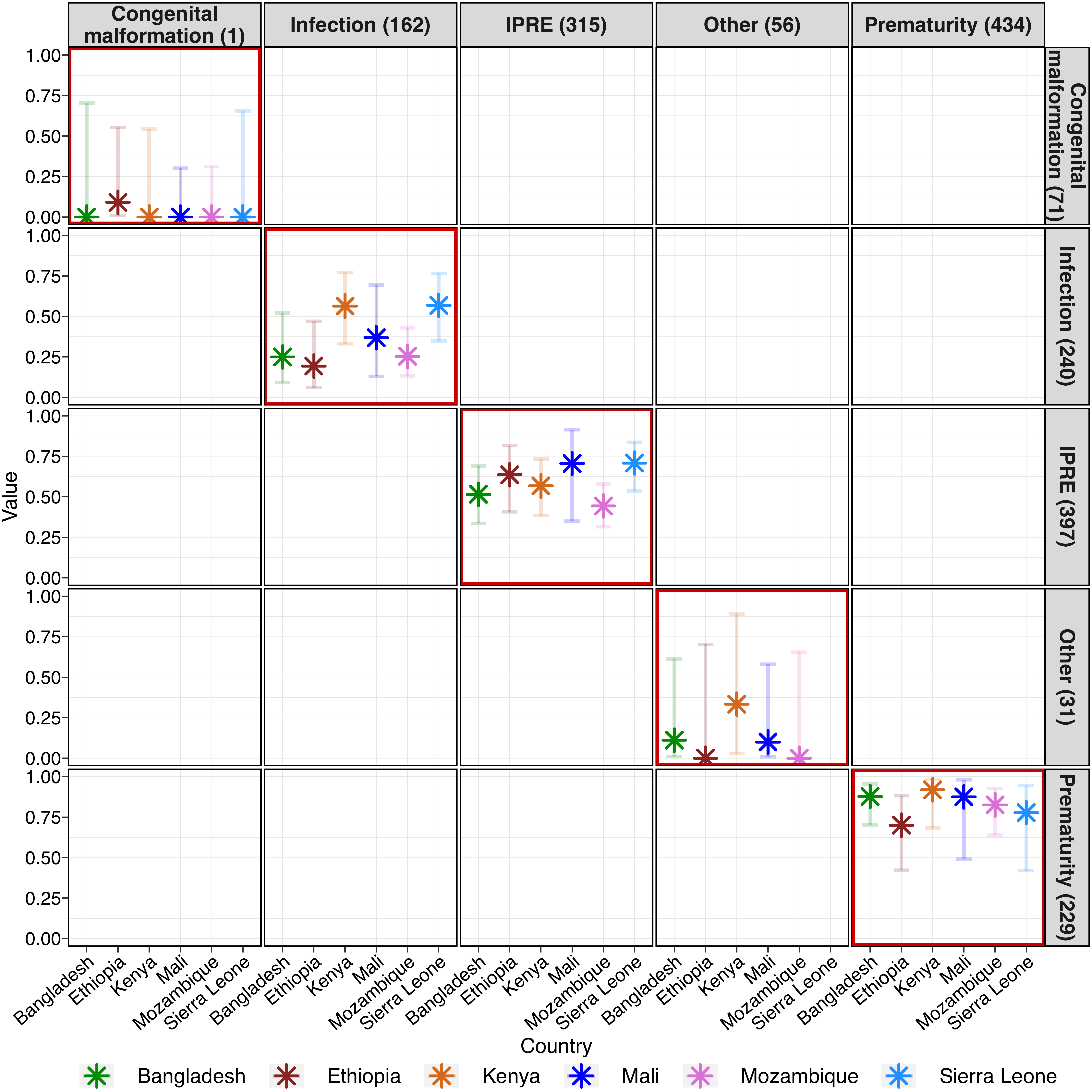}
         \caption{Sensitivity}
         \label{sfig: sensitivity}
     \end{subfigure}
     \hfill
     \begin{subfigure}[b]{0.495\textwidth}
         \centering
         \includegraphics[width=\textwidth]{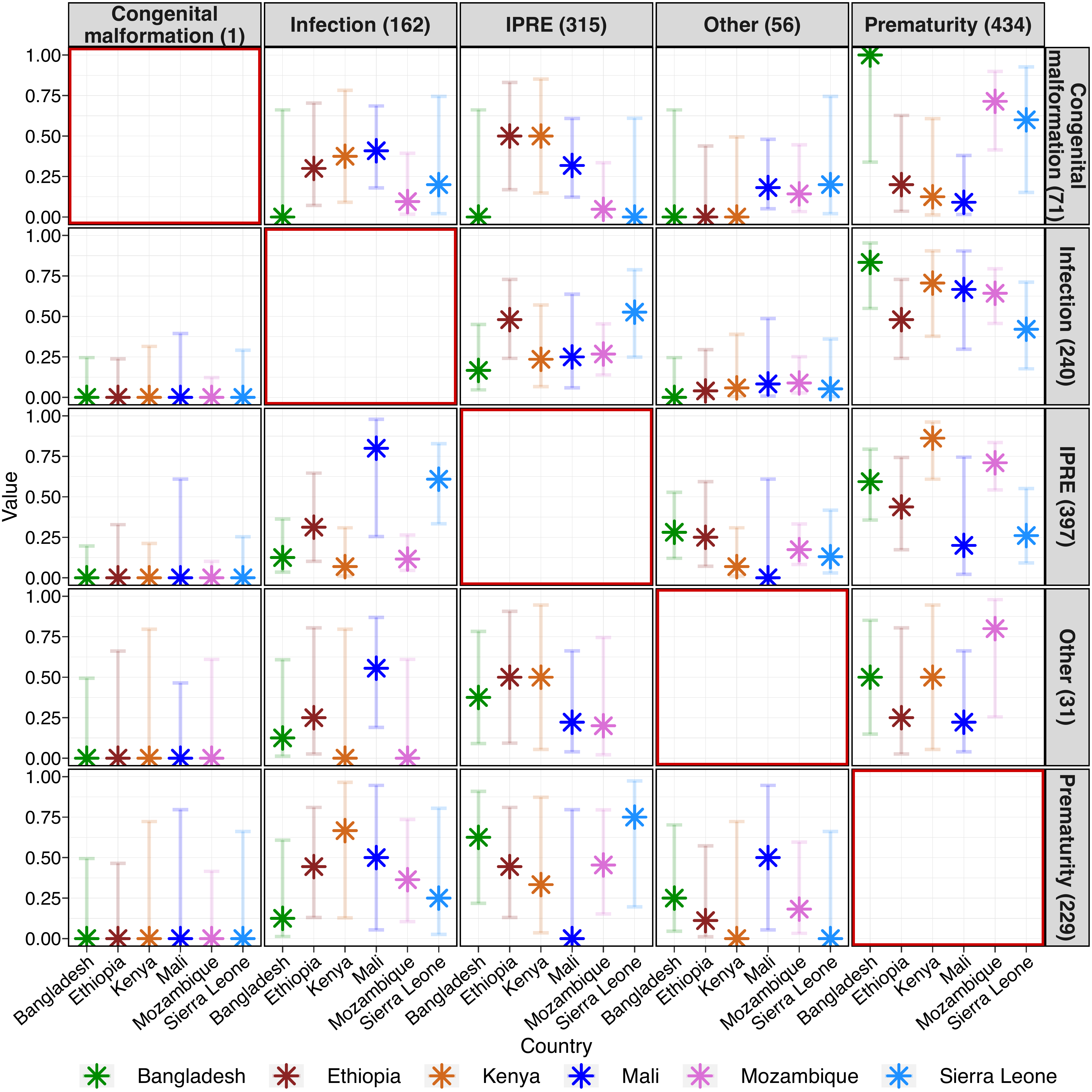}
         \caption{Relative false positive}
         \label{sfig: rfp}
     \end{subfigure}
        \caption{The exact decomposition of InsilicoVA's observed classification proportions into sensitivities and relative false positives according to decomposition (4) in the main article.}
        \label{sfig: sens rfp decomposition}
\end{figure}

\section{Default Prior Parameter Choices}\label{ssec: Default Prior Parameter Choices}

We suggest using the following parameter values for default implementation of the framework (6)--(7) in the main article: $b=d=1$ indicating the Uniform prior on intrinsic accuracies, $\bs{e} = \bs{1}$ assuming the Uniform prior on pull, and $\betaprior(0.5,0.5)$ on effect sizes (Please see Section~3.5 in the main article). Practitioners are advised to analyze the impact of prior parameter selections on their specific applications and customize them accordingly. Despite the desired continuous shrinkage, the implied marginal priors on classification rates for default choices are weakly informative, closely resembling the Jeffreys prior commonly recommended in the literature as a noninformative prior choice on Binomial and Multinomial probabilities (Please see Figure~\ref{sfig: marginal prior sensitivity&fp given pull}).
\begin{figure}[!t]
     \centering
     \begin{subfigure}[b]{0.45\textwidth}
         \centering
         \includegraphics[width=.9\textwidth]{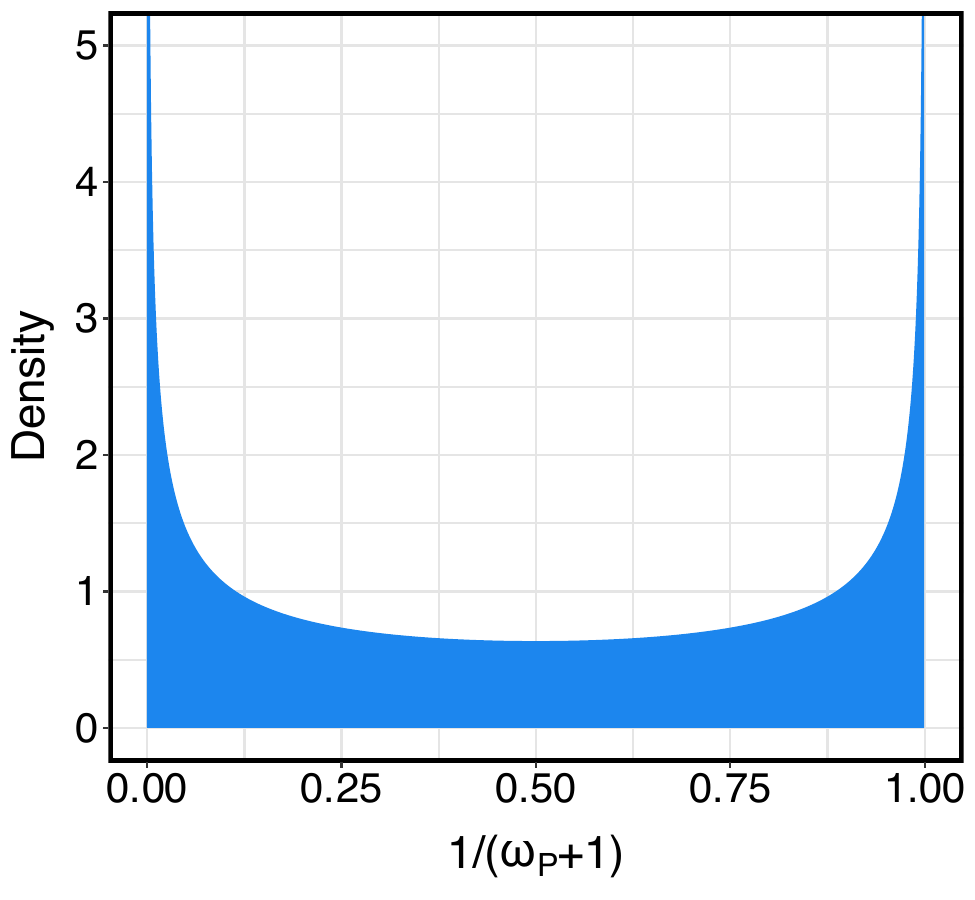}
         \caption{Transformed scale}
     \end{subfigure}
     \hfill
     \begin{subfigure}[b]{0.45\textwidth}
         \centering
         \includegraphics[width=.9\textwidth]{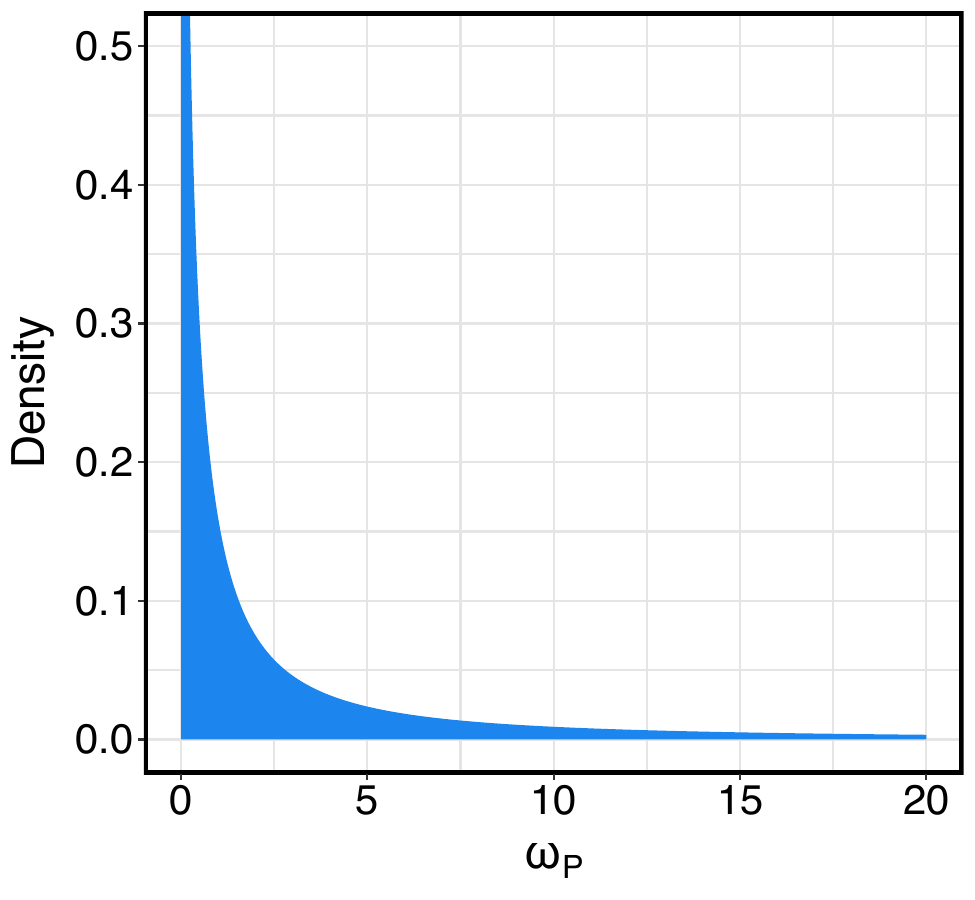}
         \caption{Original scale}
     \end{subfigure}
    \caption{(a) on the left depicts the prior on transformed effect sizes for $\varepsilon=0.5$. For the same choice, Figure (b) on the right presents the implied prior on the original scale. %\pink{AD: Use the same style of density plots for all prior figures (the latter plots shades the regions with blue.)}
        }
        \label{sfig: prior effect size}
\end{figure}
\begin{figure}[!h]
   \centering
   \includegraphics[width=.9\textwidth]{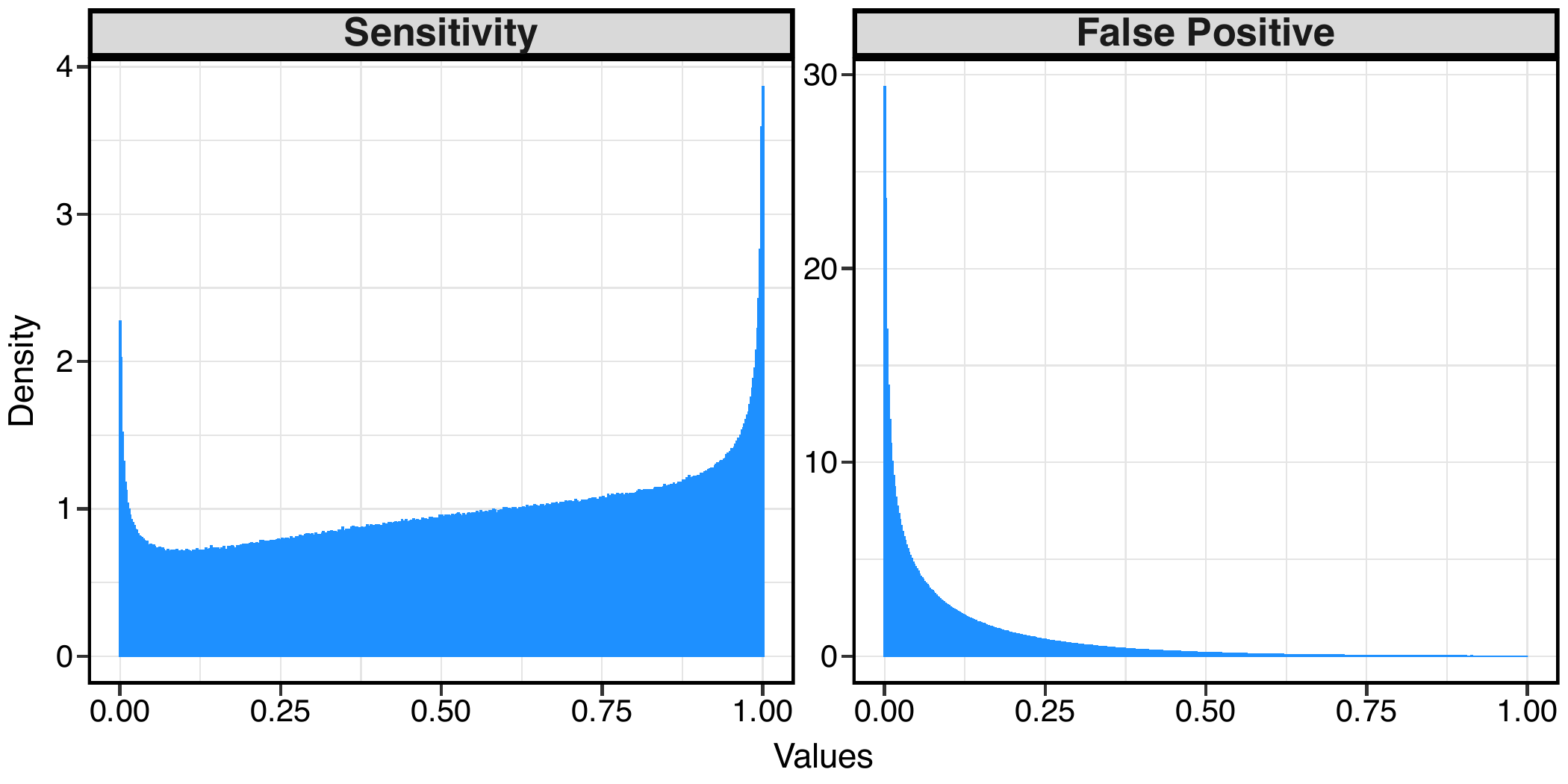}
   \caption{For 5 causes and default prior parameter choices mentioned in Section~(\ref{ssec: Default Prior Parameter Choices}), this is an example of prior on sensitivity (left panel) and false positive (right panel) in each country.}\label{sfig: marginal prior sensitivity&fp given pull}
\end{figure}

\section{Simulation Study}\label{ssec: Simulation Study}

\begin{figure}[!h]
     \centering
        \includegraphics[width=\textwidth]{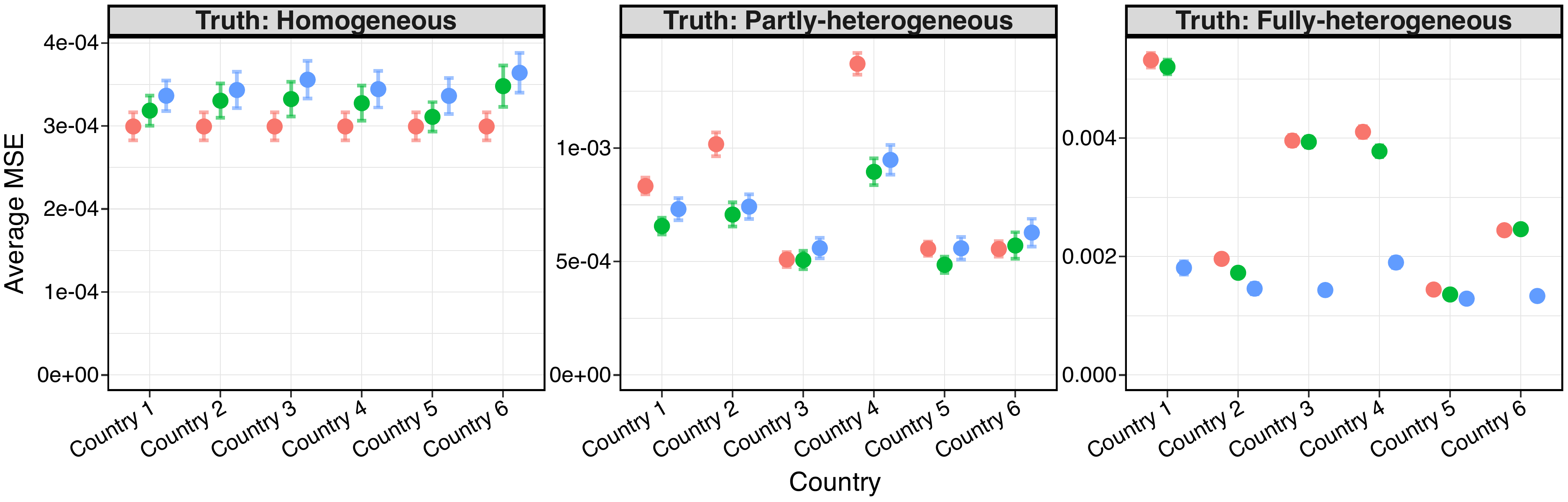}
         \includegraphics[width=\textwidth]{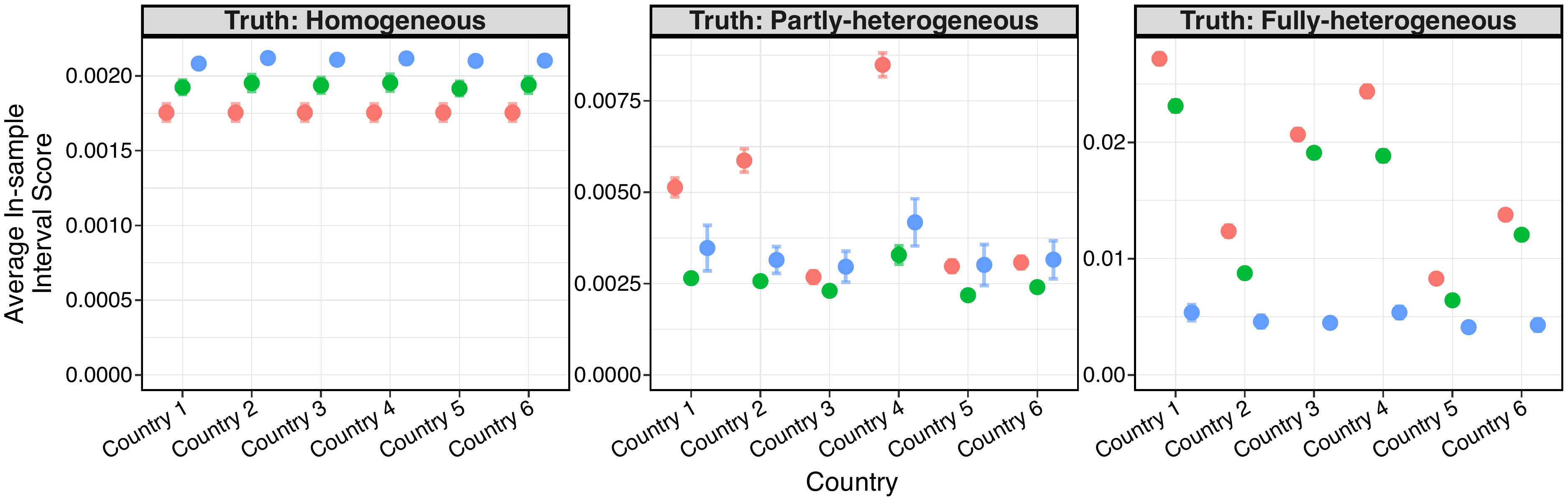}
         \includegraphics[width=\textwidth]{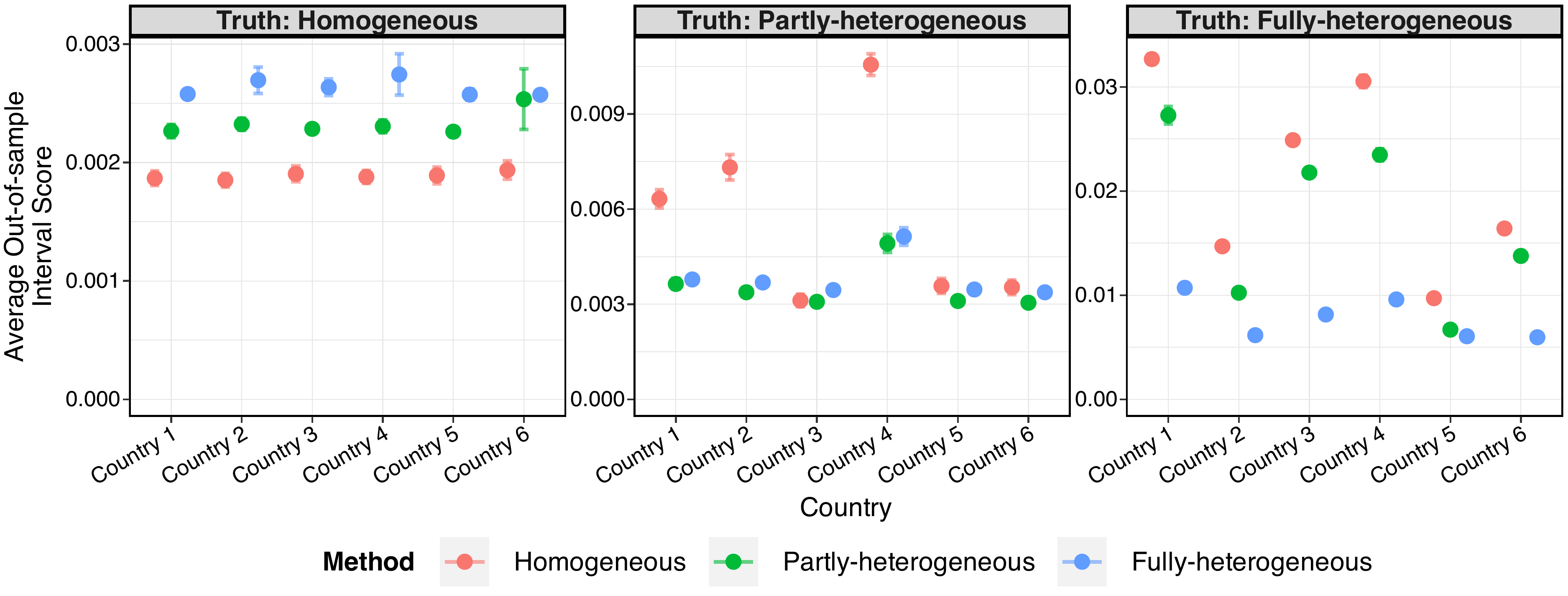}
        \caption{Averaged across all MITS-VA cause pairs over 50 replications, the figure compares in-sample mean squared errors of posterior means (top panel), in-sample interval scores of 95\% credible intervals (middle panel), and out-of-sample interval scores of 95\% prediction intervals (bottom panel) for all methods. Lower values are better. Columns show different true simulation scenarios. The error bars indicate $\pm1$ \textit{Monte-Carlo standard error}.}
        \label{sfig: simulation performance}
\end{figure}

For simulating data, we mimic features of the data observed in CHAMPS and set the number of causes $C$ to 5, the number of countries $S$ to 6, and 50 MITS cases from each country. Finally, intrinsic accuracies $(\bs{a})$, pull $(\bs{\alpha})$, and effect sizes $\omega_P$, $\omega_S$ and $\omega_R$ are set to their estimated values from the fully-heterogeneous model. Figure~9 in the main article shows that the estimate for $\omega_S$ is significantly higher than that of $\omega_R$. Consequently, this exhibits more heterogeneity in relative FP compared to sensitivities in the simulated data. Our modeling was conducted using the \texttt{Stan} software, and we employed the default prior parameter settings, as detailed in Section~\ref{ssec: Default Prior Parameter Choices}. We generated a total of 10,000 \textit{Markov Chain Monte Carlo (MCMC)} samples, discarding the initial 5,000 samples as \textit{burn-in}. The results from simulations were summarized across 50 replications.

For three methods outlined in Section~4 in the main article, Figure~\ref{sfig: simulation performance} compares their \textit{in} and \textit{out-of-sample} performances in each country averaged over MITS-VA cause pairs. The top and middle panels assess average \textit{mean square errors} of \textit{posterior means} and average \textit{interval scores} for the 95\% \textit{credible intervals}, evaluating the in-sample performance of point estimates and heterogeneity quantification \citep{Gneiting2007, scoringutils}. With the use of random effects (See (7) in the main article) in quantifying heterogeneity, the degree of heterogeneity is reflected by the uncertainty surrounding the homogeneous proportions (fixed effects). So, for analyzing the out-of-sample performance of heterogeneity prediction, we present average interval scores of 95\% \textit{prediction intervals} in the bottom panel. In summary, we find that all methods exhibit similar performance when the misclassification matrix is truly homogeneous. However, in situations where sensitivities exhibit heterogeneity, both of the heterogeneous methods perform similarly to each other, and they either surpass the performance of the homogeneous model or perform similarly to it. Finally, in cases where the misclassification matrix is fully-heterogeneous, the fully-heterogeneous method significantly outperforms both the homogeneous and partly-heterogeneous methods. This highlights the efficacy of the fully-heterogeneous method in flexibly adapting to different situations and demonstrating its superior performance in capturing and quantifying heterogeneity.

\section{Model Selection and Out-of-sample Performances in InSilicoVA's Misclassification Analysis Based on CHAMPS}

\begin{figure}[!t] %  figure placement: here, top, bottom, or page
   \centering
   \includegraphics[width=\textwidth]{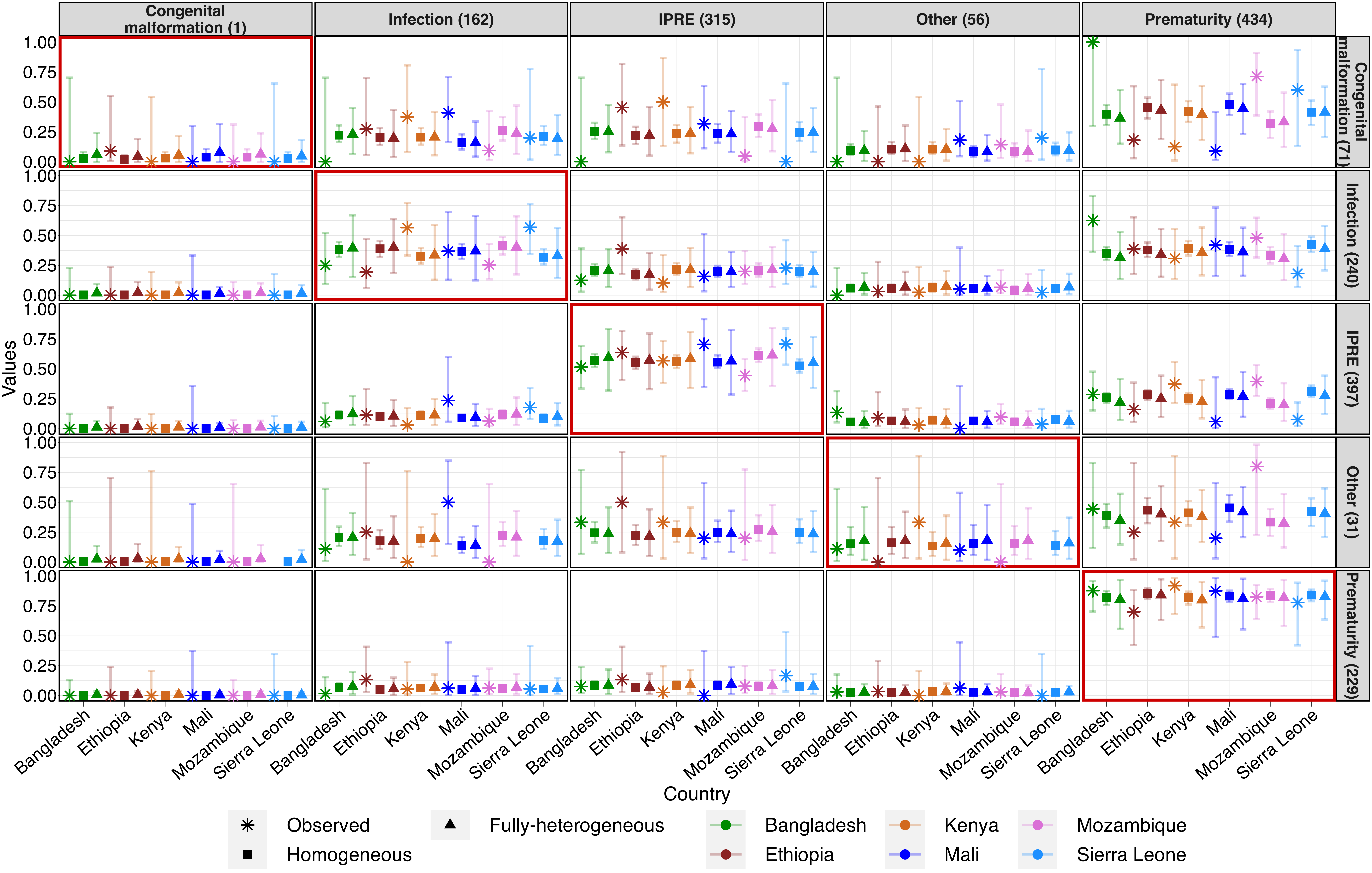}
   \caption{Comparison of observed misclassification rates with out-of-sample point (predictive mean) and uncertainty (95\% prediction interval) predictions from homogeneous and fully-heterogeneous methods. Rows and columns refer to MITS and VA-predicted causes.%\pink{AD: May need to move this to Supplement as we are half page over currently.}
   }
   \label{sfig: outofsample classification estimates}
\end{figure}
\begin{figure}[!h]
   \centering
   \includegraphics[width=.9\textwidth]{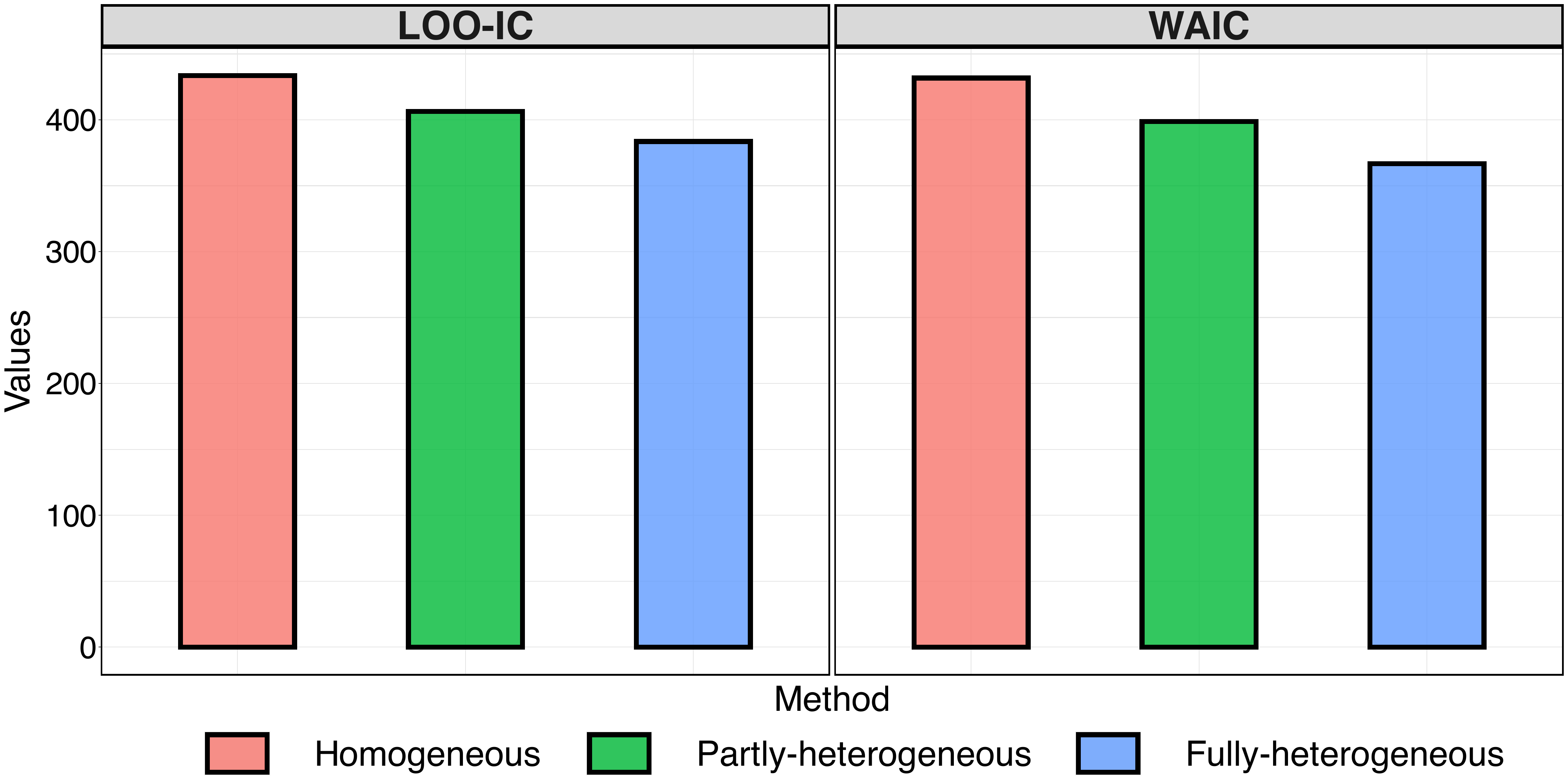}
   \caption{For neonatal deaths in CHAMPS for InSilicoVA, the bar plots compare widely applicable information criterion (WAIC) and Leave-one-out cross-validation information criterion (LOO-IC) from the three methods. Lower values are better.}\label{sfig: ic}
\end{figure}

In Figure~\ref{sfig: ic}, we compare model selection performances of different methods in the case of neonatal deaths for InSilicoVA in CHAMPS. We calculate the \textit{widely applicable information criterion (WAIC)} and \textit{Leave-one-out cross-validation information criterion (LOO-IC)}, two metrics commonly used in the literature. Compared to the homogeneous method, the partly-heterogeneous method enhances the information criteria by 6--7\%. However, the fully-heterogeneous method outperforms all, improving the criteria by 5--8\% over partly-heterogeneous and an impressive 11--15\% over the homogeneous method. This highlights the efficacy of the fully-heterogeneous method in capturing and quantifying heterogeneity.

\bibliographystyle{apalike}
\bibliography{Bibliography-MM-MC}